\documentclass[conference]{IEEEtran}
\ifCLASSINFOpdf
   \usepackage[pdftex]{graphicx}
\else
\fi

\usepackage{algorithm}
\usepackage{algpseudocode}

\usepackage{times}
\usepackage[us,12hr]{datetime}
\usepackage{enumitem}
\usepackage{subcaption}
\usepackage{comment}
\usepackage{multirow}

\usepackage{tikzpagenodes}

\IEEEoverridecommandlockouts

\begin{document}
%
\title{The Limitations of Deep Learning \\ in Adversarial Settings}

\author{\IEEEauthorblockN{}
}

\author{\IEEEauthorblockN{Nicolas Papernot\IEEEauthorrefmark{1}, Patrick McDaniel\IEEEauthorrefmark{1},
Somesh Jha\IEEEauthorrefmark{2},
Matt Fredrikson\IEEEauthorrefmark{3}, 
Z. Berkay Celik\IEEEauthorrefmark{1}, 
Ananthram Swami\IEEEauthorrefmark{4}}
\IEEEauthorblockA{\IEEEauthorrefmark{1}Department of Computer Science and Engineering, Penn State University}
\IEEEauthorblockA{\IEEEauthorrefmark{2}Computer Sciences Department, University of Wisconsin-Madison}
\IEEEauthorblockA{\IEEEauthorrefmark{3}School of Computer Science, Carnegie Mellon University}
\IEEEauthorblockA{\IEEEauthorrefmark{4}United States Army Research Laboratory, Adelphi, Maryland}
\IEEEauthorblockA{\{ngp5056,mcdaniel\}@cse.psu.edu, \{jha,mfredrik\}@cs.wisc.edu, zbc102@cse.psu.edu, ananthram.swami.civ@mail.mil}}


\maketitle

  \begin{tikzpicture}[remember picture,overlay]
    \node[align=center] at ([yshift=1em]current page text area.north) {Accepted to the 1st IEEE European Symposium on Security \& Privacy, IEEE 2016. Saarbrucken, Germany. };
  \end{tikzpicture}%

\vspace{-0.2in}



\begin{abstract}

Deep learning takes advantage of large datasets and computationally efficient training algorithms to outperform other approaches at various machine learning tasks. However, imperfections in the training phase of deep neural networks make them vulnerable to \emph{adversarial samples}: inputs crafted by adversaries with the intent of causing deep neural networks to misclassify. In this work, we formalize the space of adversaries against deep neural networks (DNNs) and introduce a novel class of algorithms to craft adversarial samples based on a precise understanding of the mapping between inputs and outputs of DNNs. In an application to computer vision, we show that our algorithms can reliably produce samples correctly classified by human subjects but misclassified in specific targets by a DNN with a 97\% adversarial success rate while only modifying on average 4.02\% of the input features per sample. We then evaluate the vulnerability of different sample classes to adversarial perturbations by defining a hardness measure. Finally, we describe preliminary work outlining defenses against adversarial samples by defining a predictive measure of distance between a benign input and a target classification. 

\end{abstract}


%
\IEEEpeerreviewmaketitle

\section{Introduction}

Large neural networks, recast as \emph{deep neural networks} (DNNs) in the mid 2000s, altered the machine learning landscape by outperforming other approaches in many tasks. This was made possible by advances that reduced the computational complexity of training~\cite{hinton2006fast}. For instance, \emph{Deep learning} (DL) can now take advantage of large datasets to achieve accuracy rates higher than previous classification techniques.  In short, DL is transforming computational processing of complex data in many domains such as vision~\cite{krizhevsky2012imagenet, taigman2014deepface}, speech recognition~\cite{dahl2012context, sainathconvolutional, sak2014long}, language processing~\cite{collobert2008unified}, financial fraud detection~\cite{PayPal}, and recently malware detection~\cite{dahl2013large}. 

This increasing use of deep learning is creating incentives for adversaries to manipulate DNNs to force misclassification of inputs. For instance, applications of deep learning use image classifiers to distinguish inappropriate from appropriate content, and text and image classifiers to differentiate between SPAM and non-SPAM email.  An adversary able to craft misclassified inputs would profit from evading detection--indeed such attacks occur today on non-DL classification systems~\cite{biggio2013evasion, biggio2014pattern, huang2011adversarial}.  In the physical domain, consider a driverless car system that uses DL to identify traffic signs~\cite{cirecsan2012multi}. If slightly altering ``STOP'' signs causes DNNs to misclassify them, the car would not stop, thus subverting the car's safety.

\begin{figure}
\centering
\includegraphics[width=0.9\columnwidth]{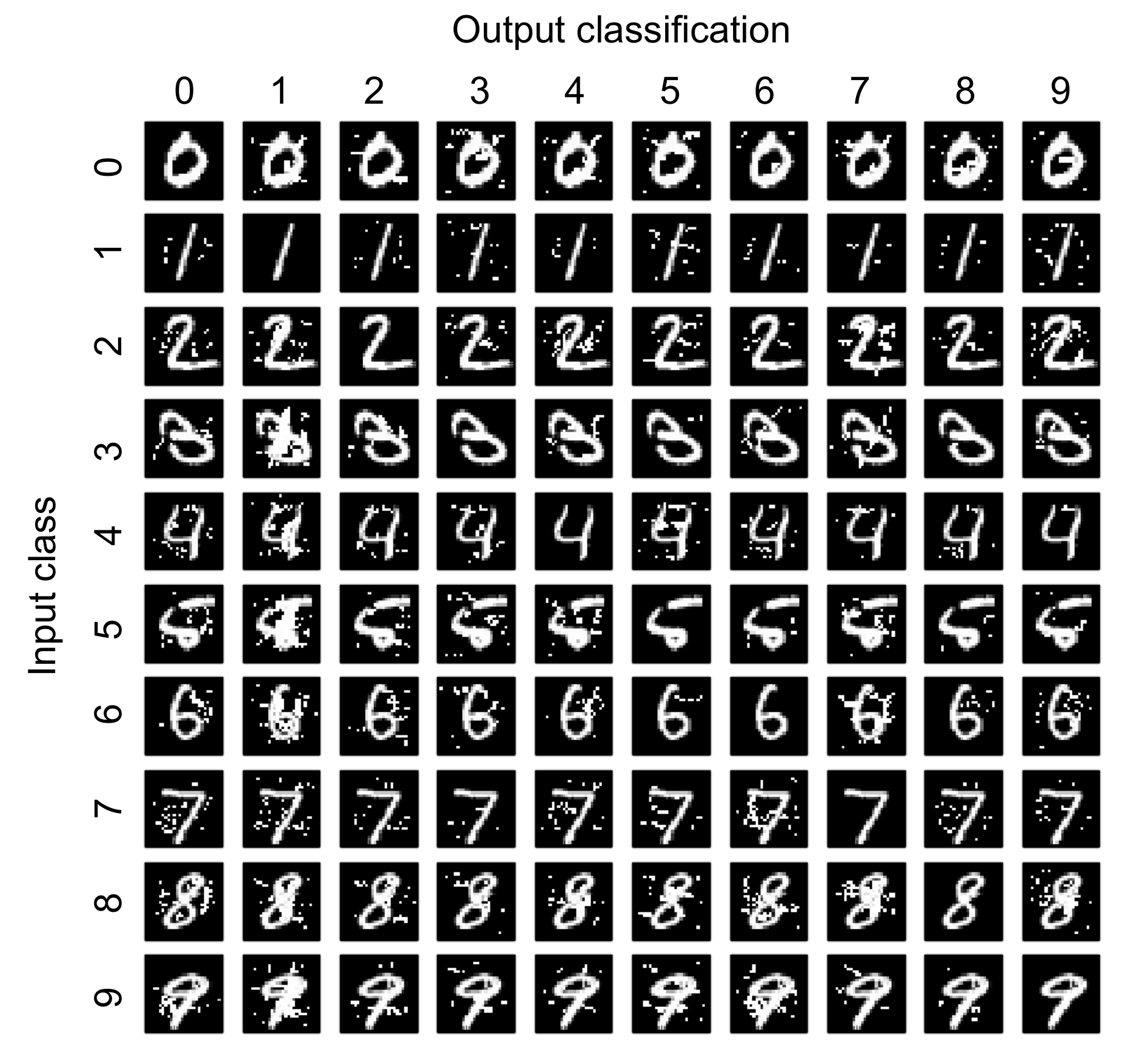}
\caption{
{\it Adversarial sample generation} - Distortion is added to input samples to force the DNN to output adversary-selected classification (min distortion $= 0.26\%$, max distortion $= 13.78\%$, and average distortion $\varepsilon=4.06\%$).  
}
\label{fig:increasing-pixels-matrix}
\end{figure}

An \emph{adversarial sample} is an input crafted to cause deep learning algorithms to misclassify. Note that adversarial samples are created at test time, after the DNN has been trained by the defender, and do not require any alteration of the training process. Figure~\ref{fig:increasing-pixels-matrix} shows examples of adversarial samples taken from our validation experiments. It shows how an image originally showing a digit can be altered to force a DNN to classify it as another digit. Adversarial samples are created from benign samples by adding distortions exploiting the imperfect generalization learned by DNNs from finite training sets~\cite{bengio2009learning}, and the underlying linearity of most components used to build DNNs~\cite{goodfellow2014explaining}. Previous work explored DNN properties that could be used to craft adversarial samples~\cite{goodfellow2014explaining, nguyen2014deep, szegedy2013intriguing}. Simply put, these techniques exploit gradients computed by network training algorithms: instead of using these gradients to update network parameters as would normally be done, gradients are used to update the original input itself, which is subsequently misclassified by DNNs. 

In this paper, we describe a new class of algorithms for adversarial sample creation against any feedforward (acyclic) DNN~\cite{rumelhart1988learning} and formalize the threat model space of deep learning with respect to the integrity of output classification.  Unlike previous approaches mentioned above, we compute a direct mapping from the input to the output to achieve an explicit adversarial goal.   Furthermore, our approach only alters a (frequently small) fraction of input features leading to reduced perturbation of the source inputs.  It also enables adversaries to apply heuristic searches to find perturbations leading to input targeted misclassifications (perturbing inputs to result in a specific output classification). 

More formally, a DNN models a multidimensional function $\mathbf{F}:\mathbf{X} \mapsto \mathbf{Y}$ where $\mathbf{X}$ is a (raw) feature vector and $\mathbf{Y}$ is an output vector. We construct an adversarial sample $\mathbf{X}^*$ from a benign sample $\mathbf{X}$ by adding a perturbation vector $\delta_\mathbf{X}$ solving the following optimization problem: \begin{equation}
\label{eq:opt-pb}
\arg \min_{\delta_\mathbf{X}} \| \delta_\mathbf{X} \| \textbf{ s.t. } \mathbf{F}\big(\mathbf{X}+\delta_\mathbf{X}\big)=\mathbf{Y}^* 
\end{equation}
where $\mathbf{X}^* = \mathbf{X}+\delta_\mathbf{X}$ is the adversarial sample, $\mathbf{Y}^*$ is the desired adversarial output, and $\| \cdot \|$ is a norm appropriate to compare the DNN inputs. Solving this problem is non-trivial, as properties of DNNs make it non-linear and non-convex~\cite{larochelle2009exploring}. Thus, we craft adversarial samples by constructing a mapping from input perturbations to output variations. Note that all research mentioned above took the opposite approach: it used output variations to find corresponding input perturbations. Our understanding of how changes made to inputs affect a DNN's output stems from the evaluation of the \emph{forward derivative}: a matrix we introduce and define as the Jacobian of the function learned by the DNN. The forward derivative is used to construct \emph{adversarial saliency maps} indicating input features to include in perturbation $\delta_\mathbf{X}$ in order to produce adversarial samples inducing a certain behavior from the DNN. 

Forward derivatives approaches are much more powerful than gradient descent techniques used in prior systems. They are applicable to both supervised and unsupervised architectures and allow adversaries to generate information for broad families of adversarial samples. Indeed, adversarial saliency maps are versatile tools based on the forward derivative and designed with adversarial goals in mind, giving greater control to adversaries with respect to the choice of perturbations.  In our work, we consider the following questions to formalize the security of DL in adversarial settings: (1) ``What is the minimal knowledge required to perform attacks against DL?'', (2) ``How can vulnerable or resistant samples be identified?'', and (3) ``How are adversarial samples perceived by humans?''. 

The adversarial sample generation algorithms are validated using the widely studied \emph{LeNet} architecture (a pioneering DNN used for hand-written digit recognition~\cite{1998gradient}) and MNIST dataset~\cite{lecun1998mnist}.
We show that any input sample can be perturbed to be misclassified as any target class with $97.10\%$ success while perturbing on average $4.02\%$ of the input features per sample. The computational costs of the sample generation are modest; samples were each generated in less than a second in our setup. Lastly, we study the impact of our algorithmic parameters on distortion and human perception of samples. This paper makes the following contributions:

\begin{itemize}

\item We formalize the space of adversaries against classification DNNs with respect to adversarial goal and capabilities. Here, we provide a better understanding of how attacker capabilities constrain attack strategies and goals. 

\item We introduce a new class of algorithms for crafting adversarial samples solely by using knowledge of the DNN architecture. These algorithms (1) exploit \emph{forward derivatives} that inform the learned behavior of DNNs, and (2) build \emph{adversarial saliency maps} enabling an efficient exploration of the adversarial-samples search space.

\item We validate the algorithms using a widely used computer vision DNN. We define and measure sample distortion and source-to-target hardness, and explore defenses against adversarial samples. We conclude by studying human perception of distorted samples.

\end{itemize}


\section{Taxonomy of Threat Models in Deep Learning}
\label{sec:taxonomy}

\begin{figure}
\centering
\includegraphics[width=\columnwidth]{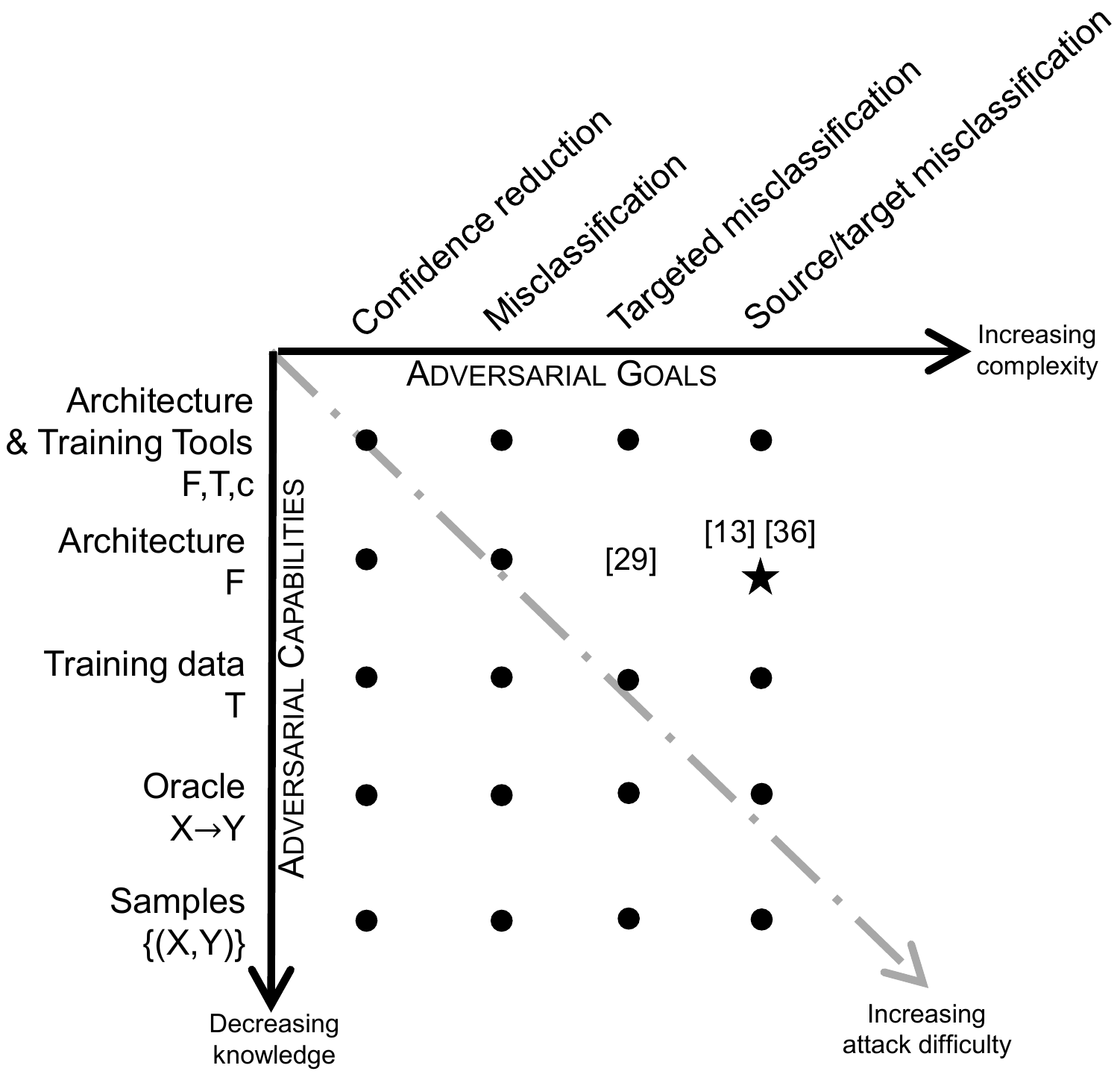}
\caption{Threat Model Taxonomy: our class of algorithms operates in the threat model indicated by a star.}
\label{fig:taxonomy-lattice}
\end{figure}

Classical threat models enumerate the goals and capabilities of adversaries in a target domain~\cite{amo94}.  This section taxonimizes threat models in deep learning systems and positions several previous works with respect to the strength of the modeled adversary.  We begin by providing an overview of deep neural networks highlighting their inputs, outputs and function. We then consider the taxonomy presented in Figure~\ref{fig:taxonomy-lattice}.

\subsection{About Deep Neural Networks}

\emph{Deep neural networks} are large neural networks organized into \emph{layers} of neurons, corresponding to successive representations of the input data. A \emph{neuron} is an individual computing unit transmitting to other neurons the result of the application of its \emph{activation function} on its input. Neurons are connected by links with different \emph{weights} and \emph{biases} characterizing the strength between neuron pairs. Weights and biases can be viewed as DNN parameters used for  information storage. We define a network \emph{architecture} to include knowledge of the network topology, neuron activation functions, as well as weight and bias values. Weights and biases are determined during \emph{training} by finding values that minimize a \emph{cost function} $c$ evaluated over the training data $T$. Network training is traditionally done by gradient descent using \emph{backpropagation}~\cite{rumelhart1988learning}. 

Deep learning can be partitioned in two categories, depending on whether DNNs are trained in a \emph{supervised} or \emph{unsupervised} manner~\cite{murphy2012machine}. Supervised training leads to models that map unseen samples using a function inferred from labeled training data. On the contrary, unsupervised training learns representations of \emph{unlabeled} training data, and resulting DNNs can be used to generate new samples, or to automate feature engineering by acting as a pre-processing layer for larger DNNs. We restrict ourselves to the problem of learning multi-class classifiers in supervised settings. These DNNs are given an input $\mathbf{X}$ and output a class probability vector $\mathbf{Y}$. Note that our work remains valid for unsupervised-trained DNNs, and leaves a detailed study of this issue for future work.

\begin{figure}
\centering
\includegraphics[width=3in]{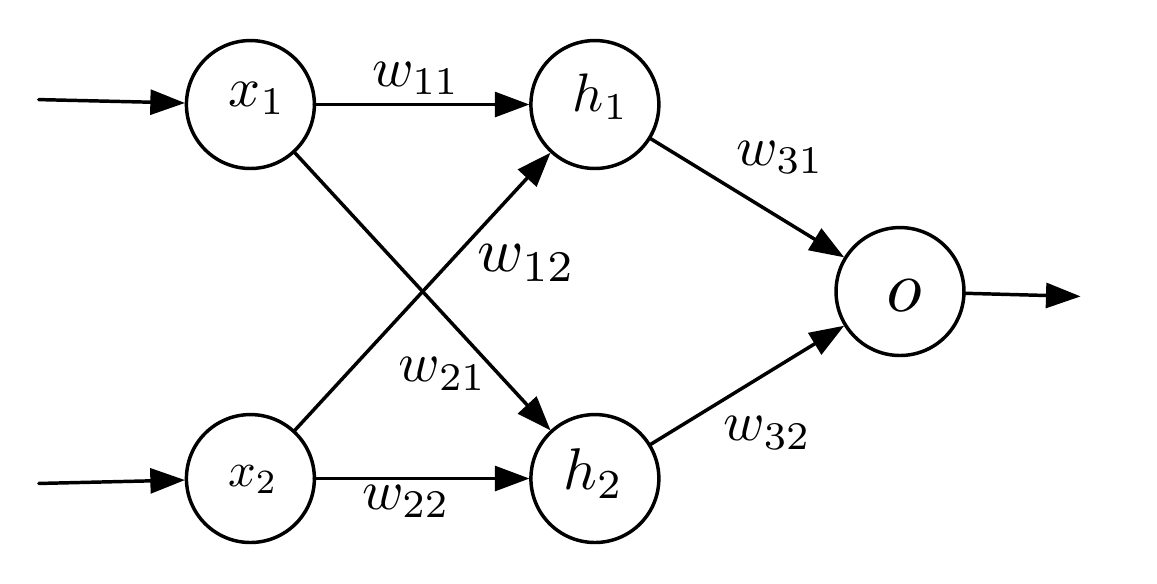}
\caption{Simplified Multi-Layer Perceptron architecture with input $\mathbf{X}=(x_1,x_2)$, hidden layer $(h_1,h_2)$, and output $o$.}
\label{fig:simplified-RBM}
\end{figure}

Figure~\ref{fig:simplified-RBM} illustrates an example shallow feedforward neural network.\footnote{A shallow neural network is a small neural network that operates (albeit at a smaller scale) identically to the DL networks considered throughout.} The network has two input neurons $x_1$ and $x_2$, a hidden layer with two neurons $h_1$ and $h_2$, and a single output neuron $o$. In other words, it is a simple multi-layer perceptron. Both input neurons $x_1$ and $x_2$ take real values in $[0,1]$ and correspond to the network input: a feature vector $\mathbf{X}=(x_1,x_2)\in [0,1]^2$. Hidden layer neurons each use the logistic sigmoid function $\phi : x \mapsto \frac{1}{1+e^{-x}}$ as their activation function. This function is frequently used in neural networks because it is continuous (and differentiable), demonstrates linear-like behavior around $0$, and saturates as the input goes to $\pm \infty$. Neurons in the hidden layers apply the sigmoid to the weighted input layer: for instance, neuron $h_1$ computes $h_1(\mathbf{X})=\phi \left( z_{h_1}(\mathbf{X})\right)$ with $z_{h_1}(\mathbf{X})=w_{11}x_1+w_{12}x_2+b_1$ where $w_{11}$ and $w_{12}$ are weights and $b_1$ a bias. Similarly, the output neuron applies the sigmoid function to the weighted output of the hidden layer where $z_{o}(\mathbf{X})=w_{31}h_1(\mathbf{X})+w_{32}h_2(\mathbf{X})+b_3$. Weight and bias values are determined during training. Thus, the overall behavior of the network learned during training can be modeled as a function: $\mathbf{F}:\mathbf{X}\rightarrow \phi \left( z_{o}(\mathbf{X})\right)$.

\subsection{Adversarial Goals}

Threats are defined with a specific function to be protected/defended.  In the case of deep learning systems, the {\it integrity} of the classification is of paramount importance.  Specifically, an adversary of a deep learning system seeks to provide an input $\mathbf{X}^*$ that results in an incorrect output classification.  The nature of the incorrectness represents the adversarial goal, as identified in the X-axis of Figure~\ref{fig:taxonomy-lattice}.  Consider four goals that impact classifier output integrity:

\begin{enumerate}

\item {\bf Confidence reduction} - reduce the output confidence classification (thereby introducing class ambiguity)

\item {\bf Misclassification} - alter the output classification to any class different from the \emph{original class}

\item {\bf Targeted misclassification} - produce inputs that force the output classification to be a specific \emph{target class}. Continuing the example illustrated in Figure~\ref{fig:increasing-pixels-matrix}, the adversary would create a set of speckles classified as a digit.

\item {\bf Source/target misclassification} - force the output classification of a specific input to be a specific \emph{target class}. Continuing the example from Figure~\ref{fig:increasing-pixels-matrix}, adversaries take an existing image of a digit and add a small number of speckles to classify the resulting image as another digit.

\end{enumerate}

The scientific community recently started exploring adversarial deep learning. Previous work on other machine learning techniques is referenced later in Section~\ref{sec:related-work}.  

Szegedy et al., introduced a system that generates
adversarial samples by perturbing inputs in a way that creates source/target
misclassifications~\cite{szegedy2013intriguing}.  The perturbations made by their work, which focused on a computer vision application, are not distinguishable by humans -- for example, small but carefully-crafted perturbations to an image of a vehicle resulted in the DNN classifying it as an ostrich. The authors named this modified input an \emph{adversarial image}, which can be generalized as part of a broader definition of \emph{adversarial samples}. When producing adversarial samples, the adversary's goal is to generate inputs that are correctly classified (or not distinguishable) by humans or other classifiers, but are misclassified by the targeted DNN.

Another example is due to Nguyen et al., who presented a method for producing images that are unrecognizable to humans, but are nonetheless labeled as recognizable objects by DNNs~\cite{nguyen2014deep}. For instance, they demonstrated how a DNN will classify a noise-filled image constructed using their technique as a television with high confidence. They named the images produced by this method \emph{fooling images}.  Here, a fooling image is one that does not have a source class but is crafted solely to perform a targeted misclassification attack.

\subsection{Adversarial Capabilities}

\newcommand{\admodel}[1]{

\vspace{3pt}
\noindent
{\bf #1 -}}

Adversaries are defined by the information and capabilities at their disposal.   
The following (and the Y-axis of Figure~\ref{fig:taxonomy-lattice}) describes a range of adversaries loosely organized by decreasing adversarial strength (and increasing attack difficulty). Note that we only considers attack conducted at test time, any tampering of the training procedure is outside the scope of this paper. 

\admodel{Training data and network architecture} This adversary has perfect knowledge of the neural network used for classification. The attacker has to access the training data $T$, functions and algorithms used for network training, and is able to extract knowledge about the DNN's architecture $\mathbf{F}$.  This includes the number and type of layers, the activation functions of neurons, as well as weight and bias matrices. He also knows which algorithm was used to train the network, including the associated loss function $c$.  This is the strongest adversary that can analyze the training data and simulate the deep neural network {\it in toto}.

\admodel{Network architecture} This adversary has knowledge of the network architecture $\mathbf{F}$ and its parameter values. For instance, this corresponds to an adversary who can collect information about both (1) the layers and activation functions used to design the neural network, and (2) the weights and biases resulting from the training phase. This gives the adversary enough information to simulate the network. Our algorithms assume this threat model, and show a new class of algorithms that generate adversarial samples for supervised and unsupervised feedforward DNNs.

\admodel{Training data}
This adversary is able to collect a \emph{surrogate} dataset, sampled from the same distribution that the original dataset used to train the DNN. However, the attacker is not aware of the architecture used to design the neural network. Thus, typical attacks conducted in this model would likely include training commonly deployed deep learning architectures using the surrogate dataset to approximate the model learned by the legitimate classifier.

\admodel{Oracle} This adversary has the ability to use the neural network (or a proxy of it) as an ``oracle''.  Here the adversary can obtain output classifications from supplied inputs (much like a chosen-plaintext attack in cryptography).  This enables differential attacks, where the adversary can observe the relationship between changes in inputs and outputs (continuing with the analogy, such as used in differential cryptanalysis) to adaptively craft adversarial samples.  This adversary can be further parameterized by the number of absolute or rate-limited input/output trials they may perform.

\admodel{Samples} This adversary has the ability to collect pairs of input and output related to the neural network classifier. However, he cannot modify these inputs to observe the difference in the output. To continue the cryptanalysis analogy, this threat model would correspond to a known plaintext attack.  These pairs are largely labeled output data, and intuition states that they would most likely only be useful in very large quantities.


\section{Approach}
\label{sec:approach}

In this section, we present a general algorithm for modifying samples so that
a DNN yields any adversarial output. We later validate this algorithm by having a classifier misclassify samples into a chosen \emph{target class}. This
algorithm captures adversaries crafting samples in the setting
corresponding to the upper right-hand corner of Figure~\ref{fig:taxonomy-lattice}. We show that
knowledge of the architecture and weight parameters\footnote{This means that the algorithm does not require  knowledge of the dataset used to train the DNN. Instead, it exploits knowledge of trained parameters.} is sufficient to derive
adversarial samples against acyclic feedforward DNNs. This requires
evaluating the DNN's \emph{forward derivative} in order to construct an
\emph{adversarial saliency map} that identifies the set of input features
relevant to the adversary's goal. Perturbing the features identified in this
way quickly leads to the desired adversarial output, for instance misclassification. Although we describe our approach with
supervised neural networks used as classifiers, it also applies to
unsupervised architectures.

\subsection{Studying a Simple Neural Network}
\label{sub:simple-nn}

\begin{figure}
\centering
\includegraphics[width=\columnwidth]{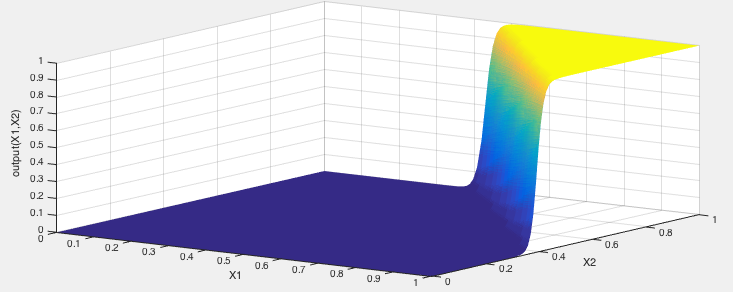}
\caption{The output surface of our simplified Multi-Layer Perceptron for the input domain $[0,1]^2$. Blue corresponds to a $0$ output while yellow corresponds to a $1$ output.}
\label{fig:simplified-RBM-output}
\end{figure}

Recall the simple architecture introduced previously in section~\ref{sec:taxonomy} and illustrated in Figure~\ref{fig:simplified-RBM}. Its low dimensionality allows us to better understand the underlying concepts behind our algorithms. We indeed show \emph{how small input perturbations found using the forward derivative can induce large variations of the neural network output}. Assuming that input biases $b_1$, $b_2$, and $b_3$ are null, we train this toy network to learn the \texttt{AND} function: the desired output is $\mathbf{F}(\mathbf{X})=x_1 \wedge x_2$ with $\mathbf{X}=(x_1,x_2)$. Note that non-integer inputs are rounded up to the closest integer, thus we have for instance $0.7 \wedge 0.3 = 0$ or $0.8 \wedge 0.6=1$.  Using backpropagation on a set of 1,000 samples, corresponding to each case of the function ($1\wedge 1 = 1$, $1\wedge 0 = 0$, $0\wedge 1 = 0$, and $0\wedge 0 = 0$), we train for 100 epochs using a learning rate $\eta=0.0663$. The overall function learned by the neural network is plotted on Figure~\ref{fig:simplified-RBM-output} for input values $\mathbf{X}\in [0,1]^2$. The horizontal axes represent the 2 input dimensions $x_1$ and $x_2$ while the vertical axis represents the network output $\mathbf{F}(\mathbf{X})$ corresponding to $\mathbf{X}=(x_1,x_2)$. 

We are now going to demonstrate how to craft adversarial samples on this neural
network. The adversary considers a \emph{legitimate} sample $\mathbf{X}$,
classified as $\mathbf{F}(\mathbf{X})=Y$ by the network, and wants to craft an
\emph{adversarial sample} $\mathbf{X}^*$ very similar to $\mathbf{X}$, but
misclassified as $\mathbf{F}(\mathbf{X}^*)=Y^* \neq Y$. Recall, that we formalized this
problem as: $$\arg \min_{\delta_\mathbf{X}} \| \delta_\mathbf{X} \| \textbf{ s.t. } \mathbf{F}\big(\mathbf{X}+\delta_\mathbf{X}\big)=\mathbf{Y}^*  $$where $\mathbf{X}^* = \mathbf{X}+\delta_\mathbf{X}$ is the adversarial sample, $\mathbf{Y}^*$ is the desired adversarial output, and $\| \cdot \|$ is a norm appropriate to compare points in the input domain. Informally, the adversary is searching for small
perturbations of the input that will incur a modification of the output into
$\mathbf{Y}^*$. Finding these perturbations can be done using optimization
techniques, simple heuristics, or even brute force. However such solutions are
hard to implement for deep neural networks because of non-convexity and
non-linearity~\cite{larochelle2009exploring}. Instead, we propose a systematic
approach stemming from the \emph{forward derivative}. 

We define the forward derivative as the Jacobian matrix of the function $\mathbf{F}$
learned by the neural network during training. For this example, the output of
$\mathbf{F}$ is one dimensional, the matrix is therefore reduced to a vector:
\begin{equation}
\label{eq:forward-derivative}
\nabla \mathbf{F}(\mathbf{X})=\left[ \frac{\partial \mathbf{F}(\mathbf{X})}{\partial x_1}, \frac{\partial \mathbf{F}(\mathbf{X})}{\partial x_2} \right]
\end{equation}
Both components of this vector are computable using the adversary's knowledge,
and later we show how to compute this term efficiently. The forward derivative for our example network is
illustrated in Figure 5, which plots the gradient for the second component
$\frac{\partial \mathbf{F}(\mathbf{X})}{\partial x_2}$ on the vertical axis against
$x_1$ and $x_2$ on the horizontal axes. We omit the plot for $\frac{\partial
\mathbf{F}(\mathbf{X})}{\partial x_1}$ because $\mathbf{F}$ is approximately symmetric on its
two inputs, making the first component redundant for our purposes. This plot
makes it easy to visualize the divide between the network's two possible
outputs in terms of values assigned to the input feature $x_2$: 0 to the left
of the spike, and 1 to its left. Notice that this aligns with
Figure~\ref{fig:simplified-RBM-output}, and gives us the information needed to
achieve our adversarial goal: find input perturbations that drive the output
closer to a desired value.

\begin{figure}
\centering
\includegraphics[width=\columnwidth]{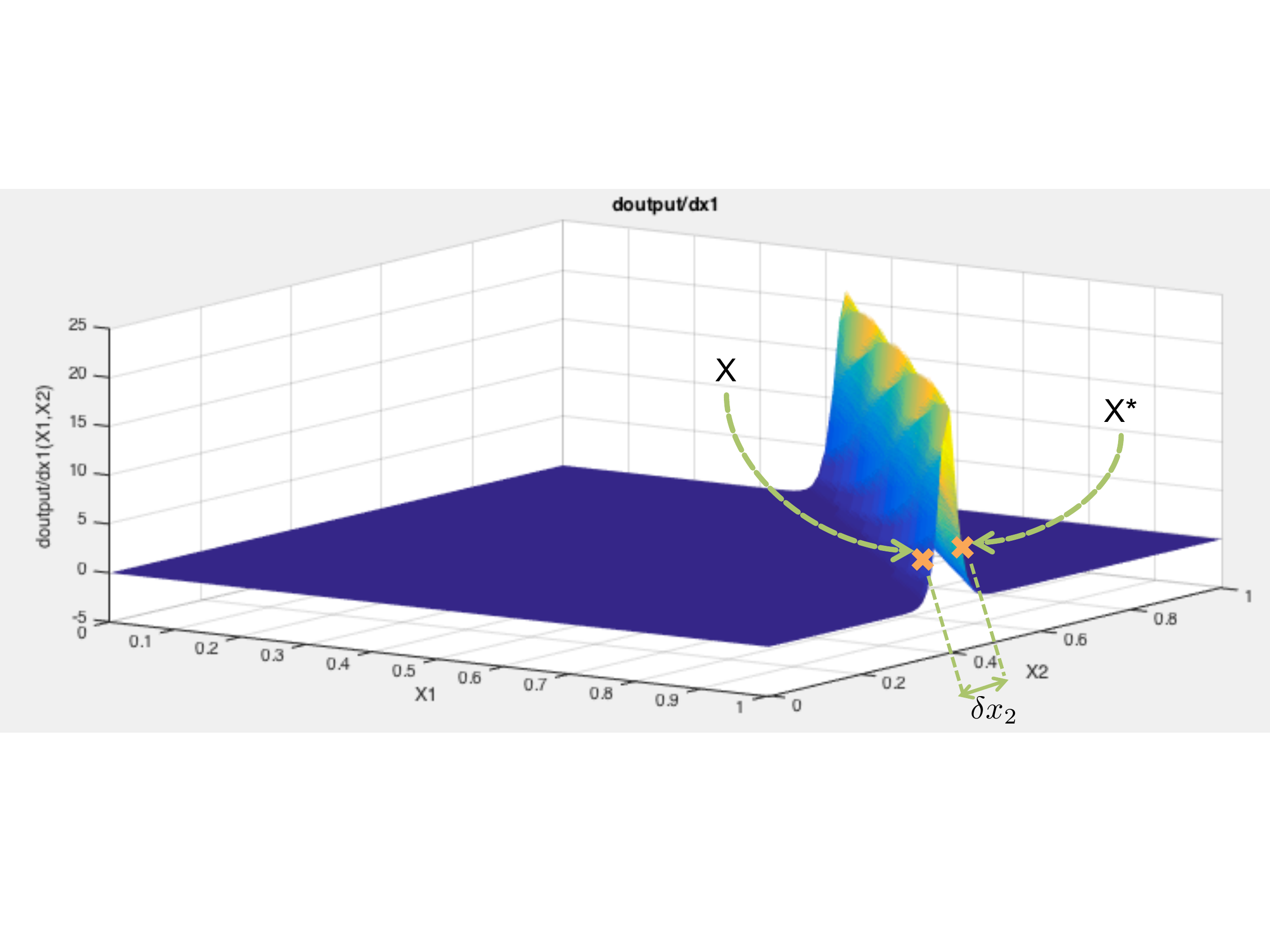}
\caption{Forward derivative of our simplified multi-layer perceptron according to input neuron $x_2$. Sample $\mathbf{X}$ is benign and $\mathbf{X}^*$ is adversarial, crafted by adding $\delta_\mathbf{X}=(0,\delta x_2)$.}
\label{fig:simplified-RBM-derivative-x2}
\end{figure}

Consulting Figure~\ref{fig:simplified-RBM-derivative-x2} alongside our example
network, we can confirm this intuition by looking at a few sample points.
Consider $\mathbf{X}=(1,0.37)$ and $\mathbf{X}^*=(1,0.43)$, which are both
located near the spike in Figure~\ref{fig:simplified-RBM-derivative-x2}.
Although they only differ by a small amount ($\delta x_2=0.05$), they cause a
significant change in the network's output, as $\mathbf{F}(\mathbf{X})=0.11$ and
$\mathbf{F}(\mathbf{X}^*)=0.95$. Recalling that we round the inputs and outputs of this
network so that it agrees with the Boolean \texttt{AND} function, we see that X*
is an adversarial sample: after rounding, $\mathbf{X}^* = (1, 0)$ and
$\mathbf{F}(\mathbf{X}^*) = 1$. Just as importantly, the forward derivative tells us
which input regions are unlikely to yield adversarial samples, and are thus more
immune to adversarial manipulations. Notice in
Figure~\ref{fig:simplified-RBM-derivative-x2} that when either input is close to
0, the forward derivative is small. This aligns with our intuition that it will
be more difficult to find adversarial samples close to $(1, 0)$ than $(1, 0.4)$.
This tells the adversary to focus on features corresponding to larger forward derivative  values in a given
input when constructing a sample, making his search more efficient and
ultimately leading to smaller overall distortions.

The takeaways of this example are thereby: \emph{(1) small input variations can lead to extreme variations of the output of the neural network, (2) not all regions from the input domain are conducive to find adversarial samples, and (3) the forward derivative reduces the adversarial-sample search space.}

\subsection{Generalizing to Feedforward Deep Neural Networks}

\begin{figure*}
\centering
\includegraphics[width=\textwidth]{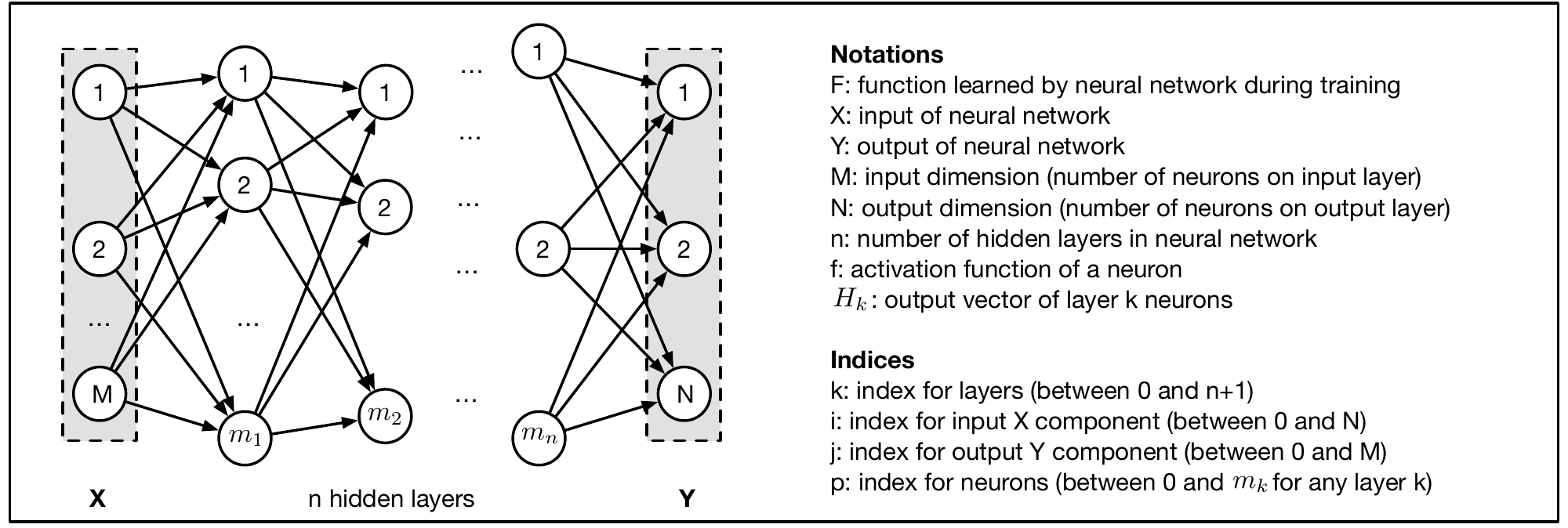}
\caption{Example architecture of a feedforward deep neural network with notations used in the paper.}
\label{fig:dnn-notations}
\end{figure*}

We now generalize this approach to any feedforward DNN, using
the same assumptions and adversary model from Section~\ref{sub:simple-nn}. The only assumptions we
make on the architecture are that its neurons form an acyclic DNN, and
each use a differentiable activation function. Note that this last assumption
is not limiting because the back-propagation algorithm imposes the same requirement. In
Figure~\ref{fig:dnn-notations}, we give an example of a feedforward deep
neural network architecture and define some notations used throughout the
remainder of the paper. Most importantly, the $N$-dimensional function
$\mathbf{F}$ learned by the DNN during training assigns an
output $\mathbf{Y}=\mathbf{F}(\mathbf{X})$ when given an $M$-dimensional input
$\mathbf{X}$. We write $n$ the number of hidden layers. Layers are indexed by $k\in 0.. n+1$
such that $k=0$ is the index of the input layer, $k\in 1..n$ corresponds to
hidden layers, and $k=n+1$ indexes the output layer.

Algorithm~\ref{alg:crafting-adversarial-samples-general} shows our process for
constructing adversarial samples. As input, the algorithm takes a benign sample
$\mathbf{X}$, a \emph{target output} $\mathbf{Y}^*$, an acyclic feedforward
DNN $\mathbf{F}$, a \emph{maximum distortion} parameter $\Upsilon$, and a
\emph{feature variation} parameter $\theta$. It returns new adversarial sample
$\mathbf{X}^*$ such that $\mathbf{F}(\mathbf{X}^*)=\mathbf{Y^*}$, and proceeds
in three basic steps: (1) compute the forward derivative $\nabla
\mathbf{F}(\mathbf{X}^*)$, (2) construct a saliency map $S$ based on the
derivative, and (3) modify an input feature $i_{\mathit{max}}$ by $\theta$. This
process is repeated until the network outputs $\mathbf{Y^*}$ or the maximum
distortion $\Upsilon$ is reached. We now detail each step.

\begin{algorithm}
\caption{\textbf{Crafting adversarial samples}\\ $\mathbf{X}$ is the benign sample, $\mathbf{Y}^*$ is the target network output, $\mathbf{F}$ is the function learned by the network during training, $\Upsilon$ is the maximum distortion, and $\theta$ is the change made to features. This algorithm is applied to a specific DNN in Algorithm~\ref{alg:crafting-adversarial-samples}.}
\label{alg:crafting-adversarial-samples-general}
\begin{algorithmic}[1]
\Require $\mathbf{X}$, $\mathbf{Y}^*$, $\mathbf{F}$, $\Upsilon$, $\theta$
\State $\mathbf{X}^*\leftarrow \mathbf{X}$
\State $\Gamma=\{1\dots|\mathbf{X}|\}$
\While{$\mathbf{F}(\mathbf{X}^*)\neq \mathbf{Y}^*$ and $||\delta_\mathbf{X}||<\Upsilon$}
	\State Compute forward derivative $\nabla \mathbf{F}(\mathbf{X}^*)$
	\State $S=\texttt{saliency\_map}\left(\nabla \mathbf{F}(\mathbf{X}^*),\Gamma,\mathbf{Y}^*\right)$
	\State Modify $\mathbf{X}^*_{i_{max}}$ by $\theta$ s.t. $i_{max}=\arg\max_{i} S(\mathbf{X},\mathbf{Y}^*)[i]$
	\State $\delta_\mathbf{X} \leftarrow \mathbf{X}^*-\mathbf{X}$
\EndWhile
\State \Return $\mathbf{X}^*$
\end{algorithmic}
\end {algorithm}

\subsubsection{Forward Derivative of a Deep Neural Network}

The first step is to compute the forward derivative for the given sample
$\mathbf{X}$. As introduced previously, this is given by:
\begin{equation}
\label{eq:forward-derivative}
\nabla \textbf{F}(\textbf{X}) = \frac{\partial \mathbf{F}(\mathbf{X})}{\partial \mathbf{X}}=\left[ \frac{\partial \mathbf{F}_j(\mathbf{X})}{\partial x_i} \right]_{i\in 1..M, j\in 1..N}
\end{equation}
This is essentially the Jacobian of the function corresponding to what the neural
network learned during training. The forward derivative computes gradients
that are similar to those computed for backpropagation, but with two
important distinctions: we take the derivative of the network directly, rather
than on its cost function, and we differentiate with respect to the input
features rather than the network parameters.
As a consequence, instead of propagating gradients backwards, we choose in our approach to
propagate them forward, as this allows us to find input
components that lead to significant changes in network outputs.

Our goal is to express $\nabla \mathbf{F}(\mathbf{X}^*)$ in terms of $\mathbf{X}$ and constant values only. To simplify our expressions, we now consider one element $(i,j)\in [1..M]\times [1..N]$ of the $M\times N$ forward derivative matrix defined in Equation~\ref{eq:forward-derivative}: that is the derivative of one output neuron $\mathbf{F}_j$ according to one input dimension $x_i$. Of course our results are true for any matrix element. We start at the first hidden layer of the neural network. We can differentiate the output of this first hidden layer in terms of the input components. We then recursively differentiate each hidden layer $k\in 2.. n$ in terms of the previous one:
\begin{equation}
\label{eq:forward-derivative-3}
\frac{\partial \mathbf{H}_k(\mathbf{X})}{\partial x_i} = \left[ \frac{\partial f_{k,p}(\mathbf{W}_{k,p}\cdot \mathbf{H}_{k-1}+b_{k,p})}{\partial x_i} \right]_{p\in 1 .. m_k}
\end{equation}
where $\mathbf{H}_k$ is the output vector of hidden layer $k$ and $f_{k,j}$ is the activation function of output neuron $j$ in layer $k$. Each neuron $p$ on a hidden or output layer indexed $k\in 1..n+1$ is connected to the previous layer $k-1$ using weights defined in vector $\mathbf{W}_{k,p}$.  By defining the weight matrix accordingly, we can define fully or sparsely connected interlayers, thus modeling a variety of architectures. Similarly, we write  $b_{k,p}$ the bias for neuron $p$ of layer $k$. By applying the chain rule, we can write a series of formulae for $k\in 2..n$: 
\begin{eqnarray}
\label{eq:forward-derivative-4}
\left.  \frac{\partial \mathbf{H}_k(\mathbf{X})}{\partial x_i} \right |_{p\in 1.. m_k} & = & \left( \mathbf{W}_{k,p}\cdot  \frac{\partial \mathbf{H}_{k-1}}{\partial x_i} \right) \times \nonumber \\ & &  \frac{\partial f_{k,p}}{\partial x_i} (\mathbf{W}_{k,p}\cdot \mathbf{H}_{k-1}+b_{k,p})
\end{eqnarray}
We are thus able to express $\frac{\partial \mathbf{H}_{n}}{\partial x_i}$. We know that output neuron $j$ computes the following expression:
$$\mathbf{F}_j(\mathbf{X})=f_{n+1,j}\left(\mathbf{W}_{n+1,j}\cdot \mathbf{H}_{n}+b_{n+1,j}\right)$$
Thus, we apply the chain rule again to obtain:
\begin{eqnarray}
\label{eq:forward-derivative-2}
\frac{\partial \mathbf{F}_j(\mathbf{X})}{\partial x_i} & = & \left( \mathbf{W}_{n+1,j}\cdot  \frac{\partial \mathbf{H}_{n}}{\partial x_i} \right) \times \nonumber \\ && \frac{\partial f_{n+1,j}}{\partial x_i} (\mathbf{W}_{n+1,j}\cdot \mathbf{H}_{n}+b_{n+1,j})
\end{eqnarray}
In this formula, according to our threat model, all terms are known but one: $\frac{\partial \mathbf{H}_{n}}{\partial x_i}$. This is precisely the term we computed recursively. By plugging these results for successive layers back in Equation~\ref{eq:forward-derivative-2}, we get an expression of component $(i,j)$ of the DNN's forward derivative. Hence, the forward derivative $\nabla \mathbf{F}$ of a network $\textbf{F}$ can be computed for any input $\textbf{X}$ by successively differentiating layers starting from the input layer until the output layer is reached. We later discuss in our methodology evaluation the computability of $\nabla \mathbf{F}$ for state-of-the-art DNN architectures. Notably, the forward derivative can be computed using symbolic differentiation. 

\subsubsection{Adversarial Saliency Maps}
We extend saliency maps previously introduced as visualization
tools~\cite{simonyan2013deep} to construct \emph{adversarial saliency maps}.
These maps indicate which input features an adversary should perturb in order
to effect the desired changes in network output most efficiently, and are thus
versatile tools that allow adversaries to generate broad classes of
adversarial samples.

\begin{figure}
\centering
\includegraphics[width=\columnwidth]{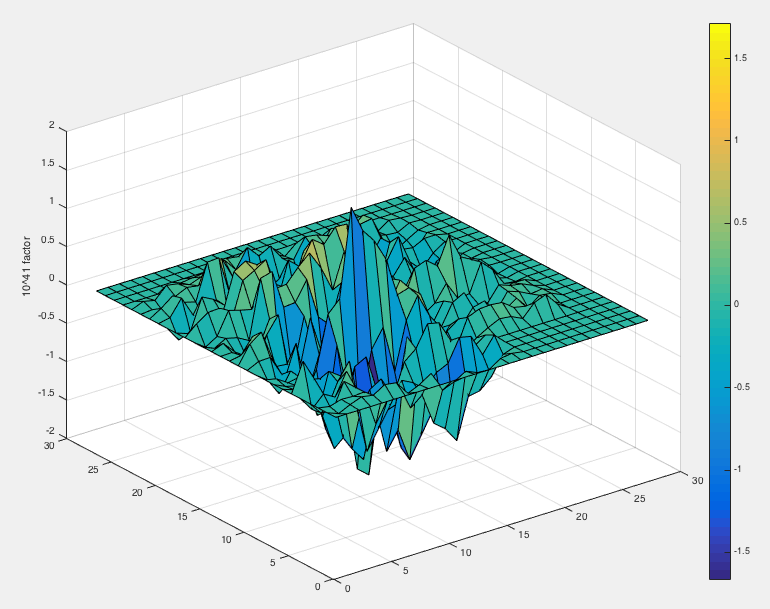}
\caption{Saliency map of a 784-dimensional input to the LeNet architecture (cf. validation section). The 784 input dimensions are arranged to correspond to the 28x28 image pixel alignment. Large absolute values correspond to features with a significant impact on the output when perturbed.}
\label{fig:saliency-map}
\end{figure}

Adversarial saliency maps are defined to suit problem-specific adversarial
goals. For instance, we later study a network used as a classifier, its output is a probability vector across classes, where
the final predicted class value corresponds to the component with the highest
probability:
\begin{equation}
\label{eq:classifier-output}
label(\mathbf{X})=\arg\max_j \mathbf{F}_j(\mathbf{X})
\end{equation} In our case, the saliency map is therefore based on the forward derivative, as
this gives the adversary the information needed to cause the neural network to
misclassify a given sample. More precisely, the adversary wants to misclassify
a sample $\mathbf{X}$ such that it is assigned a target class $t\neq
label(\mathbf{X})$. To do so, the probability of target class $t$ given by
$\mathbf{F}$, $\mathbf{F}_t(\mathbf{X})$, must be increased while the
probabilities $\mathbf{F}_j(\mathbf{X})$ of all other classes $j\neq t$
decrease, until $t = \arg\max_j \mathbf{F}_j(\mathbf{X})$. The adversary can
accomplish this by \emph{increasing} input features using the following
saliency map $S(\mathbf{X},t)$:
\begin{equation}
\label{eq:saliency-map-increasing-features}
 S(\mathbf{X},t)[i] = \left\lbrace
\begin{array}{c}
0  \mbox{ if }   \frac{\partial \mathbf{F}_{t}(\textbf{X})}{\partial \textbf{X}_i}<0  \mbox{ or } \sum_{j\neq t} \frac{\partial \mathbf{F}_{j}(\textbf{X})}{\partial \textbf{X}_i}>0\\
\left(  \frac{\partial \mathbf{F}_{t}(\textbf{X})}{\partial \textbf{X}_i}\right)  \left| \sum_{j\neq t} \frac{\partial \mathbf{F}_{j}(\textbf{X})}{\partial \textbf{X}_i}\right| \mbox{ otherwise}
\end{array}\right.
\end{equation}
where $i$ is an input feature. The condition specified on the first line
rejects input components with a negative target derivative or an overall
positive derivative on other classes. Indeed,  $\frac{\partial
\mathbf{F}_{t}(\textbf{X})}{\partial \textbf{X}_i}$ should be positive in order for
$\mathbf{F}_t(\mathbf{X})$ to increase when feature $\mathbf{X}_i$ increases. Similarly, $ \sum_{j\neq t}
\frac{\partial \mathbf{F}_{j}(\textbf{X})}{\partial \textbf{X}_i}$ needs to be negative
to decrease or stay constant when feature $\mathbf{X}_i$ is increased. The product on
the second line allows us to consider all other forward derivative components
together in such a way that we can easily compare $S(\mathbf{X},t)[i]$ for all
input features. In summary, high values of $S(\mathbf{X},t)[i]$ correspond to
input features that will either increase the target class, or decrease other
classes significantly, or both. By increasing these features, the adversary eventually misclassifies the sample into the target
class. A saliency map example is shown on Figure~\ref{fig:saliency-map}.

It is possible to define other adversarial saliency maps using the forward
derivative, and the quality of the map can have a large impact on the amount of
distortion that Algorithm~\ref{alg:crafting-adversarial-samples-general}
introduces; we will study this in more detail later. Before moving on, we
introduce an additional map that acts as a counterpart to the one given in
Equation~\ref{eq:saliency-map-increasing-features} by finding features that
the adversary should \emph{decrease} to achieve misclassification. The only
difference lies in the constraints placed on the forward derivative values and
the location of the absolute value in the second line:
\begin{equation}
\label{eq:saliency-map-decreasing-features}
 S(\mathbf{X},t)[i] = \left\lbrace
\begin{array}{c}
0  \mbox{ if }   \frac{\partial \mathbf{F}_{t}(\textbf{X})}{\partial \textbf{X}_i}>0  \mbox{ or } \sum_{j\neq t} \frac{\partial \mathbf{F}_{j}(\textbf{X})}{\partial \textbf{X}_i}<0\\
\left|  \frac{\partial \mathbf{F}_{t}(\textbf{X})}{\partial \textbf{X}_i}\right|  \left( \sum_{j\neq t} \frac{\partial \mathbf{F}_{j}(\textbf{X})}{\partial \textbf{X}_i}\right) \mbox{ otherwise}
\end{array}\right.
\end{equation}

\subsubsection{Modifying samples}

Once an input feature has been identified by an adversarial saliency map, it
needs to be perturbed to realize the adversary's goal. This is the last step
in each iteration of Algorithm~\ref{alg:crafting-adversarial-samples-general},
and the amount by which the selected feature is perturbed ($\theta$ in
Algorithm~\ref{alg:crafting-adversarial-samples-general}) is also
problem-specific. We discuss in Section~\ref{sec:validation} how this parameter should be
set in an application to computer vision. Lastly, the maximum number of
iterations, which is equivalent to the \emph{maximum distortion} allowed in a
sample, is specified by parameter $\Upsilon$. It limits the
number of features changed to craft an adversarial sample and can take any
positive integer value smaller than the number of features. Finding the
right value for $\Upsilon$ requires considering the impact of distortion on humans' perception of adversarial samples -- too much distortion might cause adversarial samples to
be easily identified by humans.

\section{Application of the Approach}
\label{sec:validation}

We formally described a class of algorithms for crafting adversarial samples misclassified by feedforward DNNs using three tools: the forward derivative, adversarial saliency maps, and the crafting algorithm. We now apply these tools to a DNN used for a computer vision classification task: handwritten digit recognition. We show that our algorithms successfully craft adversarial samples from any source class to any given target class, which for this application means that any digit can be perturbed so that it is misclassified as any other digit.

\begin{figure}
\centering
\includegraphics[width=\columnwidth]{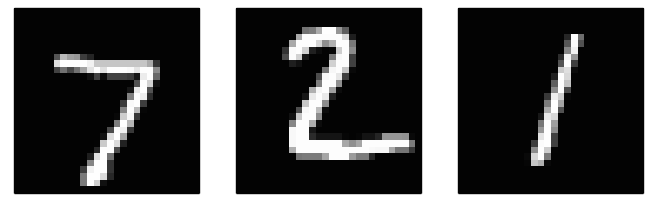}
\caption{Samples taken from the MNIST test set. The respective output vectors are: $[0,0,0,0,0,0,0.99, 0,0]$, $[0,0,0.99,0,0,0,0, 0,0]$, and $[0,0.99, 0,0,0,0,0,0,0]$, where all values smaller than $10^{-6}$ have been rounded to 0.}
\label{fig:mnist-test-set-samples}
\end{figure}

We investigate a DNN based on the well-studied LeNet architecture, which has proven to be an excellent classifier for handwritten digits~\cite{1998gradient}. Recent architectures like AlexNet~\cite{krizhevsky2012imagenet} or GoogLeNet~\cite{szegedy2014going} are heavily reliant on convolutional layers introduced in the LeNet architecture, thus making LeNet a relevant DNN to validate our approach. We have no reason to believe that our method will not perform well on larger architectures. The network input is black and white images (28x28 pixels) of handwritten digits, which are flattened as vectors of 784 features, where each feature corresponds to a pixel intensity taking normalized values between 0 and 1. This input is processed by a succession of a convolutional layer (20 then 50 kernels of 5x5 pixels) and a pooling layer (2x2 filters) repeated twice, a fully connected hidden layer (500 neurons), and an output softmax layer (10 neurons). The output is a 10 class probability vector, where each class corresponds to a digit from 0 to 9, as shown in Figure~\ref{fig:mnist-test-set-samples}. The network then labels the input image with the class assigned the maximum probability, as shown in Equation~\ref{eq:classifier-output}. We train our network using the MNIST training dataset of 60,000 samples~\cite{lecun1998mnist}.

We attempt to determine whether, using the theoretical framework introduced in previous sections, we can effectively craft adversarial samples misclassified by the DNN. For instance, if we have an image $\mathbf{X}$ of a handwritten digit 0 classified by the network as $label(\mathbf{X})=0$ and the adversary wishes to craft an adversarial sample $\mathbf{X}^*$ based on this image classified as  $label(\mathbf{X}^*)=7$, the source class is 0 and the target class is 7. Ideally, the crafting process must find the smallest perturbation $\delta_\mathbf{X}$ required to construct the adversarial sample $\mathbf{X}^*=\mathbf{X}+\delta_\mathbf{X}$. A perturbation is a set of pixel intensities -- or input feature variations -- that are added to $\mathbf{X}$ in order to craft $\mathbf{X}^*$. Note that perturbations introduced to craft adversarial samples must remain indistinguishable to humans. 

\subsection{Crafting algorithm}

Algorithm~\ref{alg:crafting-adversarial-samples} shows the crafting algorithm used in our experiments, which we implemented in Python (see Appendix~\ref{ap:validation-details} for more information regarding the implementation). It is based on Algorithm~\ref{alg:crafting-adversarial-samples-general}, but several details have been changed to accommodate our handwritten digit recognition problem. Given a network $\mathbf{F}$, Algorithm~\ref{alg:crafting-adversarial-samples} iteratively modifies a sample $\mathbf{X}$ by perturbing two input features (i.e., pixel intensities) $p_1$ and $p_2$ selected by \texttt{saliency\_map}. The saliency map is constructed and updated between each iteration of the algorithm using the DNN's forward derivative $\nabla \mathbf{F}(\mathbf{X}^*)$. The algorithm halts when one of the following conditions is met: (1) the adversarial sample is classified by the DNN with the target class $t$, (2) the maximum number of iterations $\texttt{max\_iter}$ has been reached, or (3) the feature search domain $\Gamma$ is empty. The crafting algorithm is fine-tuned by three parameters:
\begin{itemize}
\item Maximum distortion $\Upsilon$: this defines when the algorithm should stop modifying the sample in order to reach the adversarial target class. The maximum distortion, expressed as a percentage, corresponds to the maximum number of pixels to be modified when crafting the adversarial sample, and thus sets the maximum number of iterations $ \texttt{max\_iter} $ (2 pixels modified per iteration) as follows: $$ \texttt{max\_iter} = \left \lfloor \frac{784\cdot  \Upsilon}{2\cdot 100}\right \rfloor $$ where $784=28\times28$ is the number of pixels in a sample. 
\item Saliency map: subroutine $\texttt{saliency\_map}$ generates a map defining which input features will be modified at each iteration. Policies used to generate saliency maps vary with the nature of the data handled by the considered DNN, as well as the adversarial goals. We provide a subroutine example later in Algorithm~\ref{alg:increasing-saliency-map}.
\item Feature variation per iteration $\theta$: once input features have been selected using the saliency map, they must be modified. The variation $\theta$ introduced to these features is another parameter that the adversary must set, in accordance with the saliency maps she uses. 
\end{itemize}
The problem of finding good values for these parameters is a goal of our current evaluation, and is discussed later in Section~\ref{sec:evaluation}. For now, note that human perception is a limiting factor as it limits the acceptable maximum distortion and feature variation introduced. We now show the application of our framework with two different adversarial strategies. 

\begin{algorithm}
\caption{\textbf{Crafting adversarial samples for LeNet-5}\\ $\mathbf{X}$ is the benign image, $\mathbf{Y}^*$ is the target network output, $\mathbf{F}$ is the function learned by the network during training, $\Upsilon$ is the maximum distortion, and $\theta$ is the change made to pixels.}
\label{alg:crafting-adversarial-samples}
\begin{algorithmic}[1]
\Require $\mathbf{X}$, $\mathbf{Y}^*$, $\mathbf{F}$, $\Upsilon$, $\theta$
\State $\mathbf{X}^*\leftarrow \mathbf{X}$
\State $\Gamma=\{1\dots|\mathbf{X}|\}$ \Comment{search domain is all pixels}
\State $\texttt{max\_iter} = \left\lfloor \frac{784 \cdot \Upsilon}{2\cdot 100} \right\rfloor$
\State $s=\arg\max_j \mathbf{F}(\mathbf{X}^*)_j$ \Comment{source class}
\State $t=\arg\max_j \mathbf{Y}^*_j$ \Comment{target class}
\While{$s\neq t$ \& $\small{\texttt{iter}}<\small{\texttt{max\_iter}}$ \& $\Gamma\neq \emptyset$}
	\State Compute forward derivative $\nabla \mathbf{F}(\mathbf{X}^*)$
	\State $p_1,p_2=\texttt{saliency\_map}(\nabla \mathbf{F}(\mathbf{X}^*),\Gamma,\mathbf{Y}^*)$
	\State Modify $p_1$ and $p_2$ in $\mathbf{X}^*$ by $\theta$
	\State Remove $p_1$ from $\Gamma$ if $p_1==0$ or $p_1==1$
	\State Remove $p_2$ from $\Gamma$ if $p_2==0$ or $p_2==1$
	\State $s=\arg\max_j \mathbf{F}(\mathbf{X}^*)_j$
	\State $\texttt{iter}++$
\EndWhile
\State \Return $\mathbf{X}^*$
\end{algorithmic}
\end {algorithm}

\subsection{Crafting by increasing pixel intensities}

\begin{figure*}
\centering
\includegraphics[width=\textwidth]{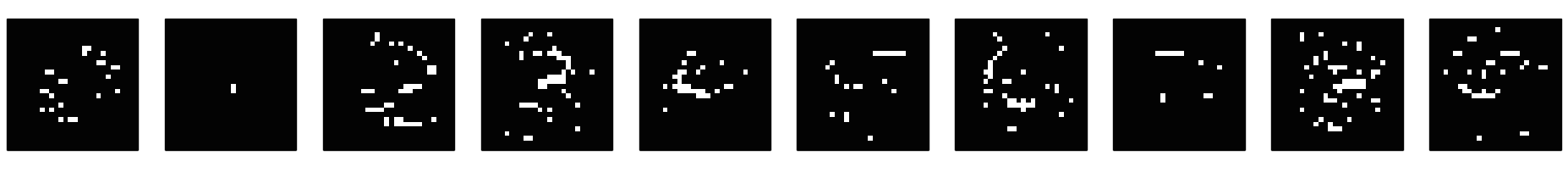}
\caption{Adversarial samples generated by feeding the crafting algorithm an empty input. Each sample produced corresponds to one target class from 0 to 9. Interestingly, for classes $0,2,3$ and $5$ one can clearly recognize the target digit.}
\label{fig:from-scratch}
\end{figure*}

The first strategy to craft adversarial samples is based on increasing the intensity of some  pixels. To achieve this purpose, we consider 10 samples of handwritten digits from the MNIST test set, one from each digit class 0 to 9. We use this small subset of samples to illustrate our techniques. We scale up the evaluation to the entire dataset in Section~\ref{sec:evaluation}. Our goal is to report whether we can reach any adversarial target class for a given source class. For instance, if we are given a handwritten 0, we increase some of the pixel intensities to produce 9 adversarial samples respectively classified in each of the classes 1 to 9. All pixel intensities changed are increased by $\theta = +1$. We discuss this choice of parameter in section~\ref{sec:evaluation}. We allow for an unlimited maximum distortion $\Upsilon=\infty$. We simply measure for each of the 90 source-target class pairs whether an adversarial sample can be produced or not.

The adversarial saliency map used in the crafting algorithm to select pixel pairs that can be increased is an application of the map introduced in the general case of classification in Equation~\ref{eq:saliency-map-increasing-features}. The map aims to find pairs of pixels $(p_1,p_2)$ using the following heuristic: 
\begin{equation}
\label{eq:saliency-increasing}
\arg\max_{(p_1,p_2)} \left(  \sum_{i= p_1,p_2}{\frac{\partial \mathbf{F}_{t}(\textbf{X})}{\partial \textbf{X}_i}}\right)  \times \left| \sum_{i= p_1,p_2}\sum_{j\neq t} \frac{\partial \mathbf{F}_{j}(\textbf{X})}{\partial \textbf{X}_i}\right|
\end{equation}
where $t$ is the index of the target class, the left operand of the multiplication operation is constrained to be positive, and the right operand of the multiplication operation is constrained to be negative. This heuristic, introduced in the previous section of this manuscript, searches for pairs of pixels producing an increase in the target class output while reducing the sum of the output of all other classes when simultaneously increased. The pseudocode of the corresponding subroutine \texttt{saliency\_map} is given in Algorithm~\ref{alg:increasing-saliency-map}.

The saliency map considers pairs of pixels and not individual pixels because selecting pixels one at a time is too strict, and very few pixels would meet the heuristic search criteria described in Equation~\ref{eq:saliency-map-increasing-features}. Searching for pairs of pixels is more likely to match the condition because one of the pixels can compensate a minor flaw of the other pixel. Let's consider a simple example: $p_1$ has a target derivative of $5$ but a sum of other classes derivatives equal to $0.1$, while $p_2$ as a target derivative equal to $-0.5$ and a sum of other classes derivatives equal to $-6$. Individually, these pixels do not match the saliency map's criteria stated in Equation~\ref{eq:saliency-map-increasing-features}, but combined, the pair does match the saliency criteria defined in Equation~\ref{eq:saliency-increasing}. One would also envision considering larger groups of input features to define saliency maps. However, this comes at a greater computational cost because more combinations need to be considered each time the group size is increased. 

In our implementation of these algorithms, we compute the forward derivative of the network using the last hidden layer instead of the output probability layer. This is justified by the extreme variations introduced by the logistic regression computed between these two layers to ensure probabilities sum up to 1, leading to extreme derivative values. This reduces the quality of information on how the neurons are activated by different inputs and causes the forward derivative to loose accuracy when generating saliency maps. Better results are achieved when working with the last hidden layer, also made up of 10 neurons, each corresponding to one digit class 0 to 9. This justifies enforcing constraints on the forward derivative. Indeed, as the output layer used for computing the forward derivative does not sum up to 1, increasing $\mathbf{F}_{t}(\textbf{X})$ does not imply that $\sum_{j\neq t} \partial \mathbf{F}_{j}(\textbf{X})$ will decrease, and vice-versa. 

\begin{algorithm}
\caption{\textbf{Increasing pixel intensities saliency map}\\ $\nabla \mathbf{F}(\mathbf{X})$ is the forward derivative, $\Gamma$ the features still in the search space, and $t$ the target class}
\label{alg:increasing-saliency-map}
\begin{algorithmic}[1]
\Require $\nabla \mathbf{F}(\mathbf{X})$, $\Gamma$, $t$
\For{each pair $(p,q) \in \Gamma$}
	\State $\alpha=\sum_{i = p,q}{\frac{\partial \mathbf{F}_{t}(\textbf{X})}{\partial \textbf{X}_i}}$
	\State $\beta=\sum_{i = p,q}\sum_{j\neq t} \frac{\partial \mathbf{F}_{j}(\textbf{X})}{\partial \textbf{X}_i}$
	\If{$\alpha>0$ and $\beta<0$ and $- \alpha\times \beta >$ max}
		\State $p_1,p_2\leftarrow p,q$
		\State $max \leftarrow - \alpha\times \beta $
	\EndIf
\EndFor
\State \Return $p_1,p_2$
\end{algorithmic}
\end {algorithm}

The algorithm is able to craft successful adversarial samples for all 90 source-target class pairs. Figure~\ref{fig:increasing-pixels-matrix} shows the 90 adversarial samples obtained as well as the 10 original samples used to craft them. The original samples are found on the diagonal. A sample on row $i$ and column $j$, when $i\neq j$, is a sample crafted from an image originally classified as source class $i$ to be misclassified as target class $j$.  

To verify the validity of our algorithms, and more specifically of our adversarial saliency maps, we run a simple experiment. We run the crafting algorithm on an empty input (all pixels initially set to an intensity of $0$) and craft one adversarial sample for each class from 0 to 9. The different samples shown in Figure~\ref{fig:from-scratch} demonstrate how adversarial saliency maps are able to identify input features relevant to classification in a class. 

\subsection{Crafting by decreasing pixel intensities}

Instead of increasing pixel intensities to achieve the adversarial targets, the second adversarial strategy decreases pixel intensities by $\theta=-1$. The implementation is identical to the exception of the adversarial saliency map. The formula is the same as previously written in Equation~\ref{eq:saliency-increasing} but the constraints are different: the left operand of the multiplication operation is now constrained to be negative, and the right operand to be positive. This heuristic, also introduced in the previous section of this paper, searches for pairs of pixels producing an increase in the target class output while reducing the sum of the output of all other classes when simultaneously decreased. 

\begin{figure}
\centering
\includegraphics[width=\columnwidth]{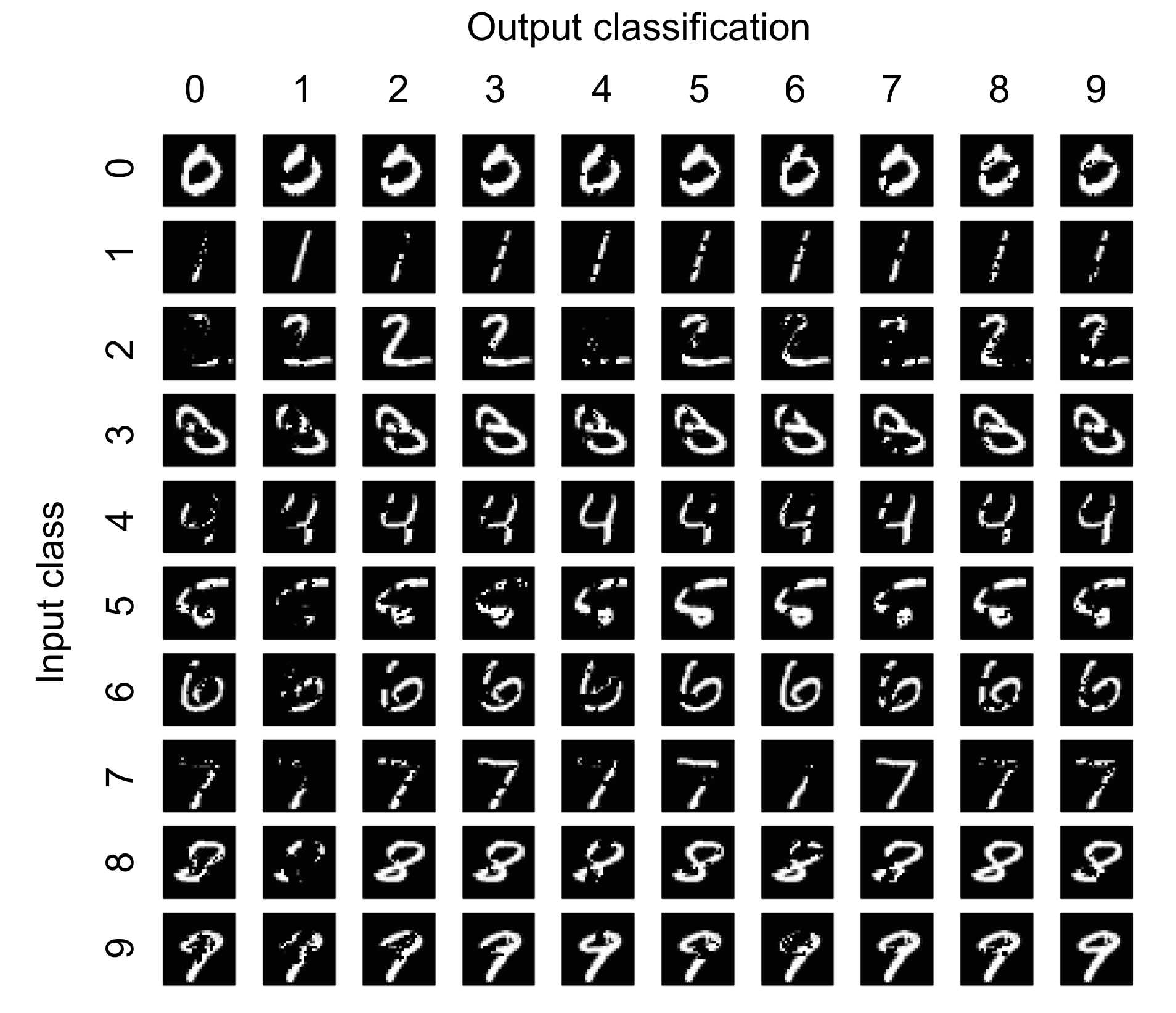}
\caption{Adversarial samples obtained by decreasing pixel intensities. Original samples from the MNIST dataset are found on the diagonal, whereas adversarial samples are all non-diagonal elements. Samples are organized by columns each corresponding to a class from 0 to 9.}
\label{fig:decreasing-pixels-matrix}
\end{figure}

The algorithm is once again able to craft successful adversarial samples for all source-target class pairs. Figure~\ref{fig:decreasing-pixels-matrix} shows the 90 adversarial samples obtained as well as the 10 original samples used to craft them. One observation to be made is that the distortion introduced by reducing pixel intensities seems harder to detect by the human eye. We address the human perception aspect with a study later in Section~\ref{sec:evaluation}. 

\section{Evaluation}
\label{sec:evaluation}

We now use our experimental setup to answer the following questions: (1) ``Can we exploit any sample?'', (2) ``How can we identify samples more vulnerable than others?'' and (3) ``How do humans perceive adversarial samples compared to DNNs?''. Our primary result is that adversarial samples can be crafted reliably for our validation problem with a $97.10\%$ success rate by modifying samples on average by $4.02\%$. We define a hardness measure to identify sample classes easier to exploit than others. This measure is necessary for designing robust defenses. We also found that humans cannot perceive the perturbation introduced to craft adversarial samples misclassified by the DNN: they still correctly classify adversarial samples crafted with a distortion smaller than $14.29\%$. 

\subsection{Crafting large amounts of adversarial samples}

Now that we previously showed the feasibility of crafting adversarial samples for all source-target class pairs, we seek to measure whether the crafting algorithm can successfully handle large quantities of distinct samples of hand-written digits. That is, we now design a set of experiments to evaluate whether or not all legitimate samples in the MNIST dataset can be exploited by an adversary to produce adversarial samples. We run our crafting algorithm on three sets of 10,000 samples each extracted from one of the three MNIST training, validation, and test subsets\footnote{Note that we extracted original samples from the dataset for convenience. Any sample can be used as an input to the adversarial crafting algorithm.}. For each of these samples, we craft 9 adversarial samples, each of them classified in one of the 9 target classes distinct from the original legitimate class. Thus, we generate 90,000 samples for each set, leading to a total of 270,000 adversarial samples. We set the maximum distortion to $\Upsilon=14.5\%$ and pixel intensities are increased by $\theta=+1$. The maximum distortion was fixed after studying the effect of increasing it on the success rate $\tau$. We found that $97.1\%$ of the adversarial samples could be crafted with a distortion of less than $14.5\%$ and observed that the success rate did not increase significantly for larger maximum distortions. Parameter $\theta$ was set to $+1$ after observing that decreasing it or giving it negative values increased the number of features modified, whereas we were interested in reducing the number of features altered during crafting. One will also notice that because features are normalized between 0 and 1, if we introduce a variation of $\theta=+1$, we always set pixels to their maximum value 1. This justifies why in Algorithm~\ref{alg:crafting-adversarial-samples}, we remove modified pixels from the search space at the end of each iteration. The impact on performance is beneficial, as we reduce the size of the feature search space at each iteration. In other words, our algorithm performs a best-first heuristic search without backtracking. 

We measure the success rate $\tau$ and distortion of adversarial samples on the
three sets of 10,000 samples. The \emph{success rate} $\tau$ is defined as the
percentage of adversarial samples that were successfully classified by the DNN
as the adversarial target class. The \emph{distortion} is defined to be the
percentage of pixels modified in the legitimate sample to obtain the adversarial
sample. In other words, it is the percentage of input features modified in order
to obtain adversarial samples. We compute two average distortion values: one
taking into account all samples and a second one only taking into account
successful samples, which we write $\varepsilon$.
Figure~\ref{fig:large-scale-sets} presents the results for the three sets from
which the original samples were extracted. The results are consistent across all
sets. On average, the success rate is $\tau = 97.10\%$, the average distortion
of all adversarial samples is $4.44\%$, and the average distortion of successful
adversarial samples is $\varepsilon = 4.02\%$. This means that the average
number of pixels modified to craft a successful adversarial sample is $32$ out
of $784$ pixels. The first distortion figure is higher because it includes
unsuccessful samples, for which the crafting algorithm used the maximum
distortion $\Upsilon$, but was unable to induce a misclassification.

\begin{figure}
\centering
\begin{tabular}{|p{1.8cm} |p{1.8cm}| p{1.8cm}| p{1.8cm}|}
\hline
 \multirow{2}{1.8cm}{Source set of $10,000$ original samples} &  \multirow{2}{1.8cm}{Adversarial samples successfully misclassified} & \multicolumn{2}{c|}{Average distortion}  \\\cline{3-4} 
   &  & All adversarial samples & Successful adversarial samples   \\ \hline
Training  & 97.05\% & 4.45\% & 4.03\% \\ \hline 
Validation  & 97.19\% & 4.41\% & 4.01\% \\ \hline 
Test & 97.05\%  & 4.45\% & 4.03\% \\ \hline 
\end{tabular}
\caption{Results on larger sets of $10,000$ samples}
\label{fig:large-scale-sets}
\end{figure}

We also studied crafting of $9,000$ adversarial samples using the decreasing saliency map. We found that the success rate $\tau=64.7\%$ was lower and the average distortion $\varepsilon=3.62\%$ slightly lower. Again, decreasing pixel intensities is less successful at producing the desired adversarial behavior than increasing pixel intensities. Intuitively, this can be understood because removing pixels reduces the information entropy, thus making it harder for DNNs to extract the information required to classify the sample. Greater absolute values of intensity variations are more confidently misclassified by the DNN.

\subsection{Quantifying hardness and building defense mechanisms}

Looking at the previous experiment, about $2.9\%$ of the $270,000$ adversarial samples were not successfully crafted. This suggests that some samples are harder to exploit than others. Furthermore, the distortion figures reported are averaged on all adversarial samples produced but not all samples require the same distortion to be misclassified. Thus, we now study the hardness of different samples in order to quantify these phenomena. Our aim is to identify which source-target class pairs are easiest to exploit, as well as similarities between distinct source-target class pairs. A class pair is a pair of a source class $s$ and a target class $t$. This hardness metric allows us to lay ground for defense mechanisms. 

\subsubsection{Class pair study}

In this experiment, we construct a deeper understanding of the crafting algorithm' success rate and average distortion for different source-target class pairs. We use the 90,000 adversarial samples crafted in the previous experiments from the 10,000 samples of the MNIST test set. 

\begin{figure}
\centering
\includegraphics[width=\columnwidth]{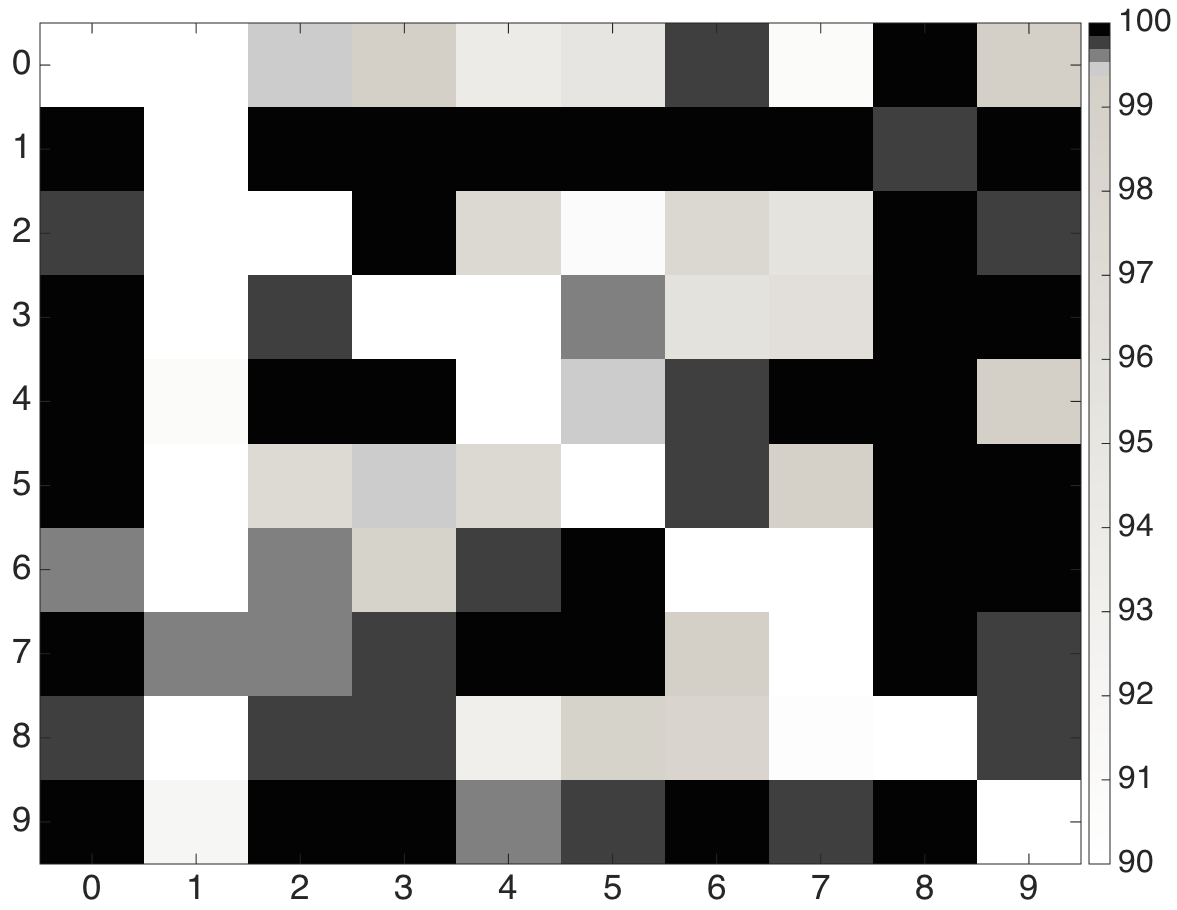}
\caption{Success rate per source-target class pair.}
\label{fig:success-matrix-test-set}
\end{figure}

We break down the success rate $\tau$ reported in Figure~\ref{fig:large-scale-sets} by source-target class pairs. This allows us to know, for a given source class, how many samples of that class were successfully misclassified in each of the target classes. In Figure~\ref{fig:success-matrix-test-set}, we draw the success rate matrix indicating which pairs are most successful. Darker shades correspond to higher success rates. The rows correspond to the success rate per source class while the columns correspond to the success rate per target class. If one reads the matrix row-wise, it can be perceived that classes 0, 2, and 8 are hard to start with, while classes 1, 7, and 9 are easy to start with. Similarly, reading the matrix column-wise, one can observe that classes 1 and 7 are very hard to make, while classes 0, 8, and 9 are easy to make.

\begin{figure}
\centering
\includegraphics[width=\columnwidth]{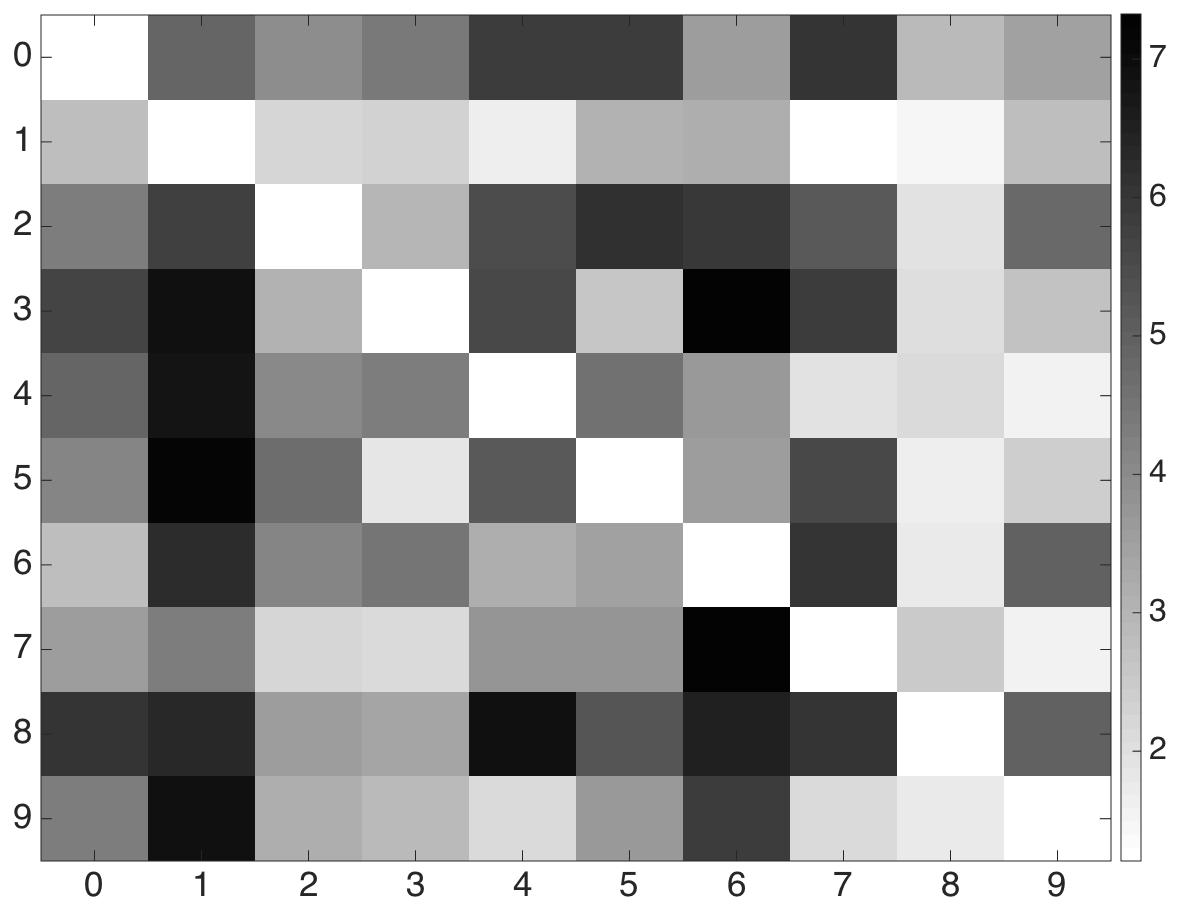}
\caption{Average distortion $\varepsilon$ of successful samples per source-target class pair. The scale is a percentage of pixels. }
\label{fig:distortion-matrix-test-set}
\end{figure}

In Figure~\ref{fig:distortion-matrix-test-set}, we report the average distortion $\varepsilon$ of successful samples by source-target class pair, thus identifying class pairs requiring the most distortion to successfully craft adversarial samples. Interestingly, classes requiring lower distortions correspond to classes with higher success rates in the previous matrix. For instance, the column corresponding to class 1 is associated with the highest distortions, and it was the column with the least success rates in the previous matrix. Indeed, the higher the average distortion of a class pair is, the more likely samples in that class pair are to reach the maximum distortion, and thus produce unsuccessful adversarial samples. 

To better understand why some class pairs were harder to exploit, we tracked the evolution of class probabilities during the crafting process. We observed that the distortion required to leave the source class was higher for class pairs with high distortions whereas the distortion required to reach the target class, once the source class had been left, remained  similar. This correlates with the fact that some source classes are more confidently classified by the DNN then others. 

\subsubsection{Hardness measure}

\begin{figure}
\centering
\includegraphics[width=\columnwidth]{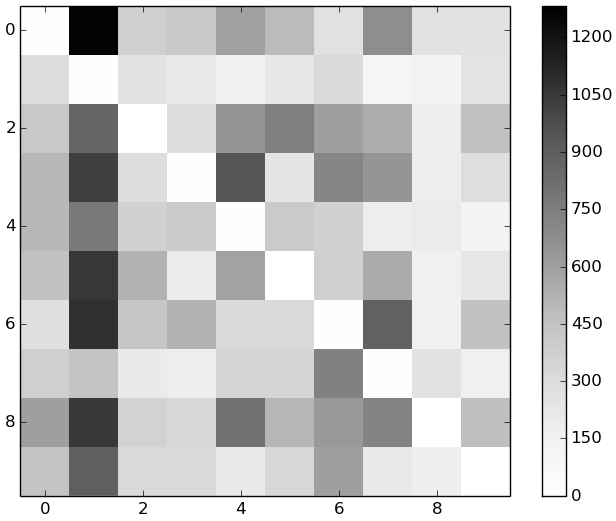}
\caption{Hardness matrix of source-target class pairs. Darker shades correspond to  harder to achieve misclassifications.}
\label{fig:hardness-class-pairs-matrix}
\end{figure}

Results indicating that some source-target class pairs are not as easy as others lead us to question the existence of a measure quantifying the distance between two classes. This is relevant to a defender seeking to identify which classes of a DNN are most vulnerable to adversaries. We name this measure the \emph{hardness} of a target class relatively to a given source class. It normalizes the average distortion of a class pair $(s,t)$ relatively to its success rate:
\begin{equation}
\label{hardness-class-pair}
H(s,t)=\int_{\tau} \varepsilon(s,t,\tau) \mathrm{d}\tau
\end{equation} 
where $\varepsilon(s,t,\tau)$ is the average distortion of a set of samples for the corresponding success rate $\tau$. In practice, these two quantities are computed over a finite number of samples by fixing a set of $K$ maximum distortion parameter values $\Upsilon_k$ in the crafting algorithm where $k\in 1..K$. The set of maximum distortions gives a series of pairs $(\varepsilon_k, \tau_k)$ for $k\in 1 .. K$. Thus, the practical formula used to compute the hardness of a source-destination class pair can be derived from the trapezoidal rule:
\begin{equation}
\label{hardness-class-pair}
H(s,t)\approx  \sum_{k=1}^{K-1} \left( \tau_{k+1} - \tau_k\right) \frac{\varepsilon(s,t,\tau_{k+1})+ \varepsilon(s,t,\tau_k)}{2}
\end{equation} 
We computed the hardness values for all classes using a set of $K=9$ maximum distortion  values $\Upsilon\in \{ 0.3,1.3,2.6,5.1,7.7,10.2,12.8,25.5,38.3 \}\%$ in the algorithm. Average distortions $\varepsilon$ and success rates $\tau$ are averaged over 9,000 adversarial samples for each maximum distortion value $\Upsilon$. Figure~\ref{fig:hardness-class-pairs-matrix} shows the hardness values $H(s,t)$ for all pairs $(s,t)\in \{0..9\}^2$. The reader will observe that the matrix has a shape similar to the average distortion matrix plotted on Figure~\ref{fig:distortion-matrix-test-set}. However, the hardness measure is more accurate because it is plotted using a series of maximum distortions. 

\subsubsection{Adversarial distance}
 
The measure introduced lays ground towards finding defenses against adversarial samples. Indeed, if the hardness measure were to be predictive instead of being computed after adversarial crafting, the defender could identify vulnerable inputs. Furthermore, a predictive measure applicable to a single sample would allow a defender to evaluate the vulnerability of specific samples as well as class pairs. We investigated several complex estimators including convolutional transformations of the forward derivative or Hessian matrices. However, we found that simply using a formulae derived from the intuition behind adversarial saliency maps gave enough accuracy for predicting the hardness of samples in our experimental setup.

\begin{figure}
\centering
\includegraphics[width=\columnwidth]{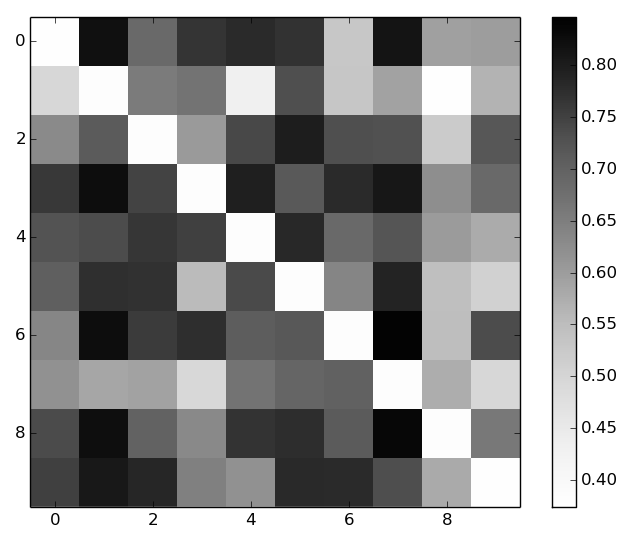}
\caption{Adversarial distance averaged per source-destination class pairs computed with 1000 samples.}
\label{sumofnonzerosaliencymap}
\end{figure}

We name this predictive measure the \emph{adversarial distance} of sample $\mathbf{X}$ to class $t$ and write it $A(\mathbf{X},t)$. Simply put, it estimates the distance between a sample $\mathbf{X}$ and a target class $t$. We define the distance as:
\begin{equation}
\label{eq:adversarial-distance}
A(\mathbf{X},t)=1-\frac{1}{M} \sum_{i\in0..M} 1_{S(\mathbf{X},t)[i] > 0}
\end{equation}

where $1_{E}$ is the indicator function for event $E$ (i.e., is $1$ if and only if $E$ is true). In a nutshell, $A(\mathbf{X},t)$ is the normalized number of non-zero elements in the adversarial saliency map of $\mathbf{X}$ computed during the first crafting iteration in Algorithm~\ref{alg:crafting-adversarial-samples}. The closer the adversarial distance is to 1, the more likely sample $\mathbf{X}$ is going to be harder to misclassify in target class $t$. Figure~\ref{sumofnonzerosaliencymap} confirms that this formulae is empirically well-founded. It illustrates the value of the adversarial distance averaged per source-destination class pairs, making it easy to compare the average value with the hardness matrix computed previously after crafting samples. To compute it, we slightly altered Equation~\ref{eq:adversarial-distance} to sum over pairs of features, reflecting the observations made during our validation process. 

This notion of distance between classes intuitively defines a metric
for the \emph{robustness} of a network $\mathbf{F}$ against adversarial
perturbations. We suggest the following definition :
\begin{equation}
\label{eq:robustness}
R(\mathbf{F})=\min_{(\mathbf{X},t)} A(\mathbf{X},t)
\end{equation}
where the set of samples $\mathbf{X}$ considered is sufficiently large to represent the input domain of the network. A good approximation of the robustness can be computed  with the training dataset. Note that the $\min$ operator used here can be replaced by other relevant operators, like the statistical expectation. The study of various operators is left as future work.

\subsection{Study of human perception of adversarial samples}

Recall that adversarial samples must not only be misclassified as the target
class by deep neural networks, but also visually appear (be classified) as the source class by humans. 
To evaluate this property, we ran an experiment using 349 human
participants on the Mechanical Turk online service. We presented three original
or adversarially altered samples from the MNIST dataset to human participants. 
To paraphrase, participants were asked for each sample: (a) `is this sample
a numeric digit?', and (b) `if yes to (a) what digit is it?'.  These two
questions were designed to determine how distortion and intensity rates effected
human perception of the samples.

The first experiment was designed to identify a baseline perception
rate for the input data.  The $74$ participants were presented 3 of 222
unaltered samples randomly picked from the original MNIST data set. Respondents identified $97.4\%$ as digits and classified the digits correctly $95.3\%$ of the samples.

\begin{figure}[t!]
\centering
\includegraphics[width=\columnwidth]{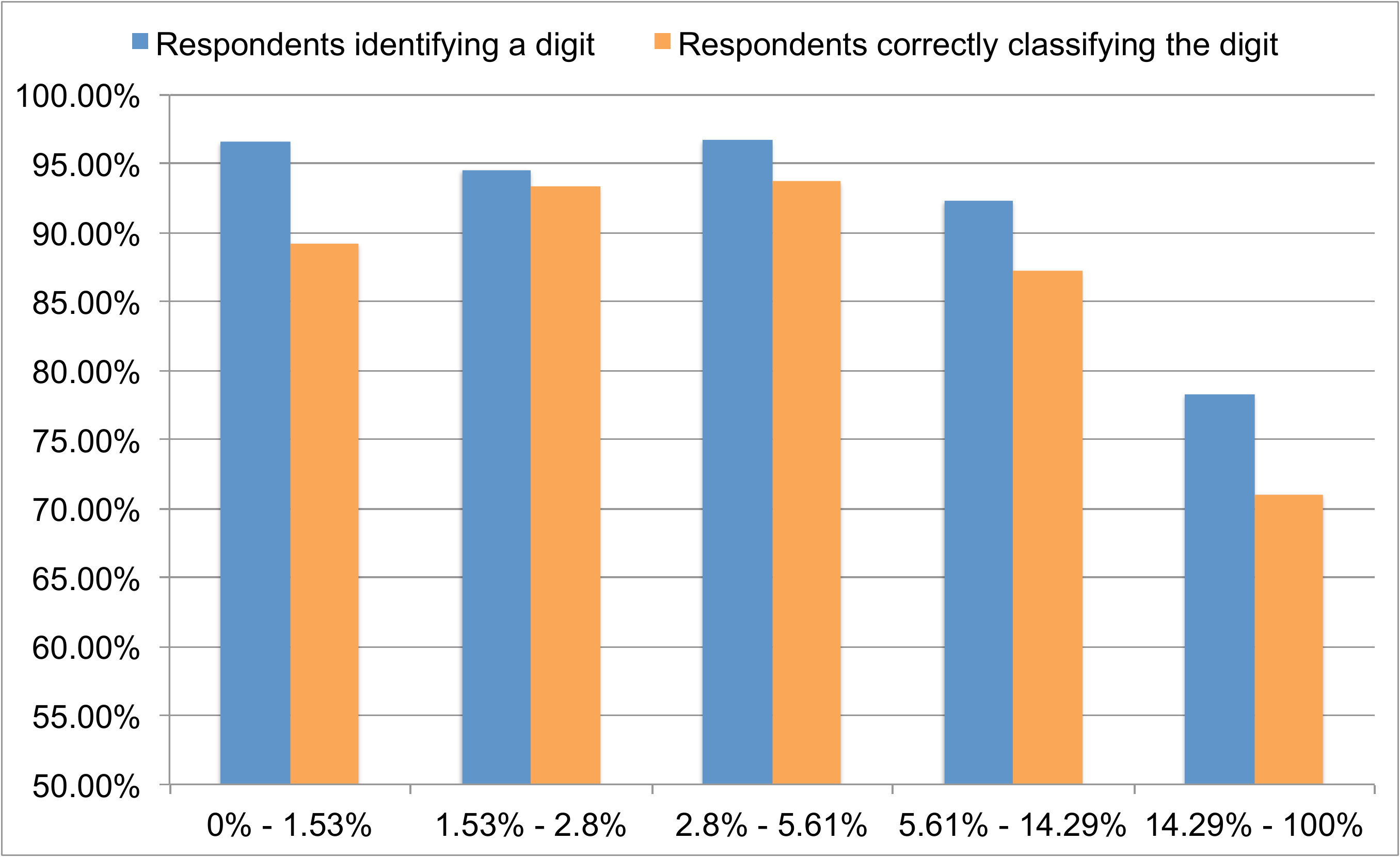}
\caption{Human perception of different distortions $\varepsilon$.}
\label{fig:mturk-distortion}
\end{figure}

\begin{figure}[t!]
\centering
\includegraphics[width=\columnwidth]{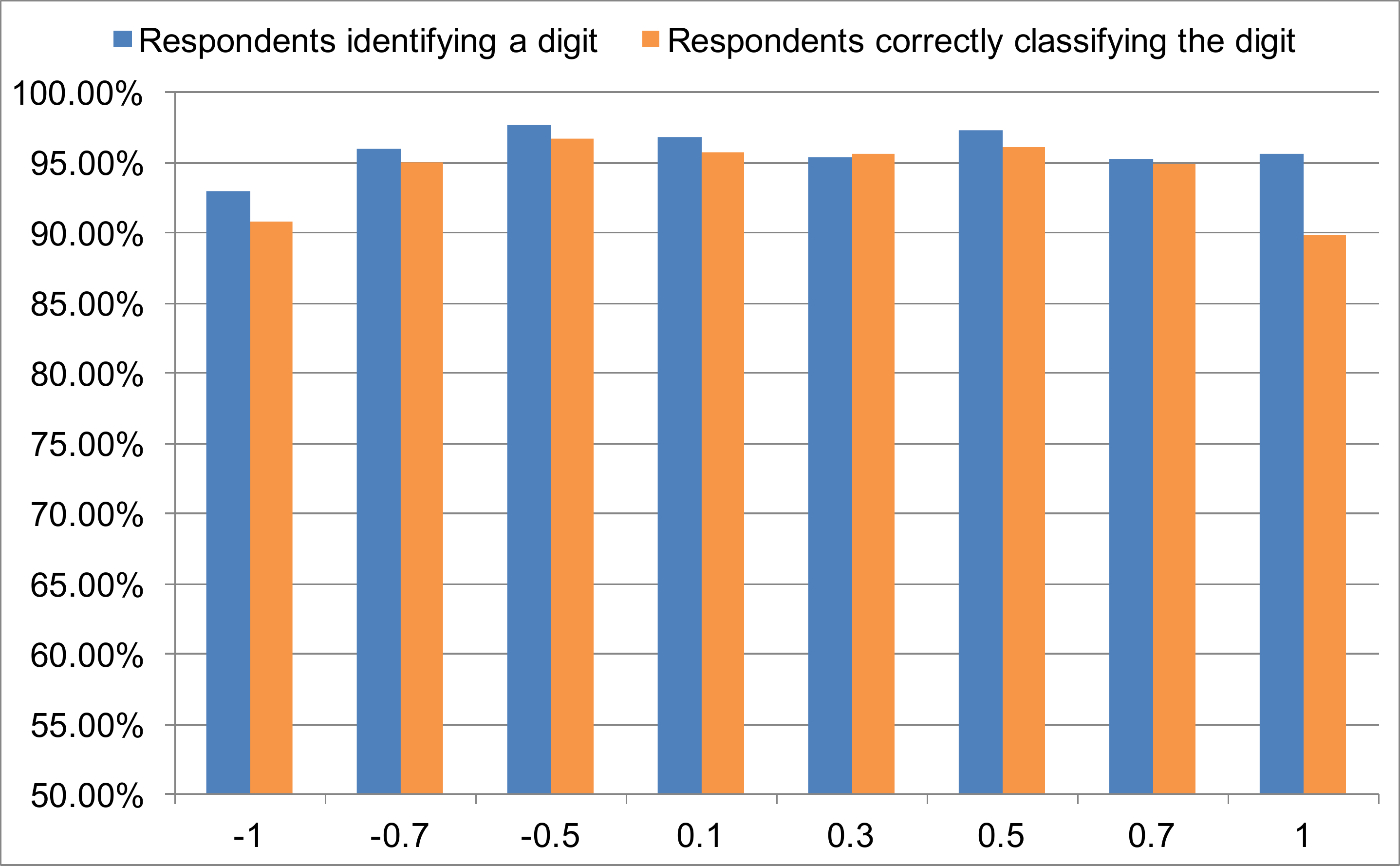}
\caption{Human perception of different intensity variations $\theta$.}
\label{fig:mturk-intensity-variation}
\end{figure} 

Shown in Figure~\ref{fig:mturk-distortion}, a second set of experiments
attempted to evaluate how the amount of distortion ($\varepsilon$) impacts human
perception.  Here, $184$ participants were presented with a total of $1707$ samples
with varying levels of distortion (and features altered with an intensity increase $\theta=+1$).  The experiments
showed that below a threshold ($\varepsilon=14.29\%$ distortion), participants were able to
identify samples as digits ($95\%$) and correctly classify them ($90\%$) only
slightly less accurately than the unaltered samples.  The classification rate
dropped dramatically ($71\%$) at distortion rates above the threshold.

A final set of experiments evaluate the impact of intensity variations
($\theta$) on perception, as shown Figure~\ref{fig:mturk-intensity-variation}. 
The $203$ participants were accurate at identifying $5,355$ samples as digits
($96\%$) and classifying them correctly ($95\%$).  At higher absolute intensities
($\theta = -1$ and $\theta = +1$), specific digit classification decreased slightly ($90.5\%$ and $90\%$), but identification as digits was largely unchanged.

%
While preliminary, these experiments confirm that the overwhelming number of
generated samples retain human recognizability.  Note that because we can
generate samples with less than the distortion threshold for the almost all of
the input data, ($\varepsilon \le 14.29\%$ for roughly 97\% in the MNIST data), we
can produce adversarial samples that humans will mis-interpret---thus meeting
our adversarial goal.  Furthermore, altering feature distortion intensity
provides even better results: at $-0.7\leq\theta \le +0.7$, humans classified the sample
data at essentially the same rates as the original sample data.


\section{Discussion}
\label{sec:discussion}

We introduced a new class of algorithms that systematically craft adversarial
samples misclassified by a DNN once an adversary possesses knowledge of the
DNN architecture. Although we focused our work on DL techniques
used in the context of classification and trained with supervised methods,
our approach is also applicable to unsupervised architectures.
Instead of achieving a given target class, the adversary achieves a target
output $\mathbf{Y^*}$. Because the output space is more complex, it might be
harder or impossible to match $\mathbf{Y^*}$. In that case, Equation~\ref{eq:opt-pb} would need to be relaxed with an acceptable distance between
the network output $\mathbf{F}(\mathbf{X}^*)$ and the adversarial target
$\mathbf{Y^*}$. Thus, the only remaining assumption made in this
paper is that DNNs are feedforward. In other words, we did not
consider recurrent neural networks, with cycles in their architecture, as
the forward derivative must be adapted to accommodate such networks.

One of our key results is reducing the distortion---the number of features
altered---to craft adversarial samples, compared to previous work. We believe this 
makes adversarial crafting much easier for input 
domains like malware executables, which are not as easy to perturb as images~\cite{biggio2014poisoning,fogla2006evading}. This distortion reduction comes with a performance cost. Indeed, more elaborate
but accurate saliency map formulae are more expensive to compute
for the attacker. We would
like to emphasize that our method's high success rate can be further
improved by adversaries only interested in crafting a limited number of samples. Indeed,
to lower the distortion of one particular sample, an adversary can use adversarial saliency maps to fine-tune the perturbation introduced. On the other hand, if an adversary wants to craft large
amounts of adversarial samples, performance is important. In our evaluation, we
balanced these factors to craft adversarial samples against the DNN
in less than a second. As far as our algorithm implementation was
concerned, the most computationally expensive steps were the matrix
manipulations required to construct adversarial saliency maps from the forward
derivative matrix. The complexity is dependent of the number of input features.
These matrix operations can be made more efficient, notably by making better use of
GPU-accelerated computations.

Our efforts so far represent a first but meaningful step towards mitigating
adversarial samples: the hardness and adversarial distance metrics lay out bases for defense mechanisms. Although designing such defenses is outside of the scope of this paper, we outline two classes of defenses: (1) 
 adversarial sample detection and (2) improvements of DNN robustness.  
 
Developing techniques for adversarial sample detection is a reactive solution. During our
experimental process, we noticed that adversarial samples can for instance be
detected by evaluating the regularity of samples. More specifically, in our
application example, the sum of the squared difference between
each pair of neighboring pixels is always higher for adversarial
samples than for benign samples. However, there is no a priori reason to
assume that this technique will reliably detect adversarial samples in different
settings, so extending this approach is one avenue for future work. Another
approach was proposed in~\cite{gu2014towards},
but it is unsuccessful as by stacking the denoising auto-encoder used for
detection with the original DNN, the adversary can again produce adversarial
samples.

The second class of solutions seeks to improve training to in return
increase the robustness of DNNs. Interestingly, the problem of adversarial
samples is closely linked to training. Work on generative adversarial networks
showed that a two player game between two DNNs can lead to the
generation of new samples from a training set~\cite{goodfellow2014generative}.
This can help augment training datasets. Furthermore, adding
adversarial samples to the training set can act like a
regularizer~\cite{goodfellow2014explaining}. We also observed in our experiments
that training with adversarial samples makes crafting additional adversarial
samples harder. Indeed, by adding 18,000
adversarial samples to the original MNIST training dataset, we trained a new
instance of our DNN. We then run our algorithms again on this newly trained
network and crafted a set of 9,000 adversarial samples. Preliminary analysis of
these adversarial samples crafted showed that the success rate was reduced by $7.2\%$
while the average distortion increased by $37.5\%$, suggesting that training
with adversarial samples can make DNNs more robust.


\section{Related Work}
\label{sec:related-work}

The security of machine learning~\cite{barreno2010security} is an active research topic within the security and machine learning communities. A broad taxonomy of attacks and required adversarial capabilties are discussed in~\cite{huang2011adversarial} and~\cite{ barreno2006can} along with considerations for building defense mechanisms. Biggio et al. studied classifiers in adversarial settings and outlined a framework securing them~\cite{biggio2014security}. However, their work does not consider DNNs but rather other techniques used for binary classification like logistic regression or Support Vector Machines. Generally speaking, attacks against machine learning can be separated into two categories, depending on whether they are executed during training~\cite{biggio2011support} or at test time~\cite{biggio2012poisoning}.

Prior work on adversarial sample crafting against DNNs derived a simple technique corresponding to the \emph{Architecture and Training Tools} threat model, based on the backpropagation procedure used during network training~\cite{goodfellow2014explaining, nguyen2014deep, szegedy2013intriguing}. This approach creates adversarial samples by defining an optimization problem based on the DNN's cost function. In other words, instead of computing gradients to update DNN weights, one computes gradients to update the input, which is then misclassified as the target class by a DNN. The alternative approach proposed in this paper is to identify input regions that are most relevant to its classification by a DNN. This is accomplished by computing the saliency map of a given input, as described by Simonyan et al. in the case of DNNs handling images~\cite{simonyan2013deep}. We extended this concept to create adversarial saliency maps highlighting regions of the input that need to be perturbed in order to accomplish the adversarial goal.

Previous work by Yosinki et al. investigated how features are transferable between deep neural networks~\cite{yosinski2014transferable}, while Szegedy et al. showed that adversarial samples can indeed be misclassified across models~\cite{szegedy2013intriguing}. They report that  once an adversarial sample is generated for a given neural network architecture, it is also likely to be misclassified in neural networks designed differently, which explains why the attack is successful. However, the effectiveness of this kind of attack depends on (1) the quality and size of the surrogate dataset collected by the adversary, and (2) the adequateness of the adversarial network used to craft adversarial samples.  


\section{Conclusions}
\label{sec:conclusion}

Broadly speaking, this paper has explored adversarial behavior in deep learning systems. In addition to exploring the goals and capabilities of DNN adversaries, we introduced a new class of algorithms to craft adversarial samples based on computing \emph{forward derivatives}. This technique allows an adversary with knowledge of the network architecture to construct \emph{adversarial saliency maps} that identify features of the input that most significantly impact output classification. These algorithms can reliably produce samples correctly classified by human subjects but misclassified in specific targets by a DNN with a 97\% adversarial success rate while only modifying on average 4.02\% of the input features per sample.  

Solutions to defend DNNs against adversaries can be divided in two classes: detecting adversarial samples and improving the training phase. The detection of adversarial samples remains an open problem.  Interestingly, the universal approximation theorem formulated by Hornik et al. states one hidden layer is sufficient to represent arbitrarily accurately a function~\cite{hornik1989multilayer}.  Thus, one can intuitively conceive that improving the training phase is key to resisting adversarial samples.  

In future work, we plan to address the limitations of DNN trained in an unsupervised manner as well as cyclical recurrent neural networks (as opposed to acyclical networks considered throughout this paper). Also, as most models of our taxonomy have yet to be researched, this leaves room for further investigation of DL in various adversarial settings. 


\section*{Acknowledgment}

The authors would like to warmly thank Dr. Damien Octeau and Aline Papernot for insightful discussions about this work. Research was sponsored by the Army Research Laboratory and was accomplished
under Cooperative Agreement Number W911NF-13-2-0045 (ARL Cyber Security
CRA). The views and conclusions contained in this document are those of the
authors and should not be interpreted as representing the official policies,
either expressed or implied, of the Army Research Laboratory or the U.S.
Government. The U.S. Government is authorized to reproduce and distribute
reprints for Government purposes notwithstanding any copyright notation here
on.



%

\bibliographystyle{abbrv}
\bibliography{initial-biblio}

\appendix

\subsection{Validation setup details}
\label{ap:validation-details}

To train and use the deep neural network, we use Theano~\cite{bergstra2010theano}, a Python package designed to simplify large-scale scientific computing. Theano allows us to efficiently implement the network architecture, the training through back-propagation, and the forward derivative computation. We configure Theano to make computations with float32 precision, because they can then be accelerated using graphics processors. Indeed, all our experiments are facilitated using GPU acceleration on a machine equipped with a Xeon E5-2680 v3 processor and a Nvidia Tesla K5200 graphics processor. 

Our deep neural network makes some simplifications, suggested in the Theano Documentation~\cite{TheanoTutorial}, to the original LeNet-5 architecture. Nevertheless, once trained on batches of $500$ samples taken from the MNIST dataset~\cite{lecun1998mnist} with a learning parameter of $\eta=0.1$ for $200$ epochs, the learned network parameters exhibits a $98.93\%$ accuracy rate on the MNIST training set and $99.41\%$ accuracy rate on the MNIST test set, which are comparable to state-of-the-art accuracies. 

\end{document}